\documentclass[useAMS,usenatbib]{mn2e}
\usepackage{graphicx}
\usepackage{rotating}
\usepackage{lscape}
\usepackage{makeidx}
\usepackage{listings}
\usepackage{longtable}
\usepackage{array}

\title[Excess {[O{\sc{ii}}]} emission in radio-loud quasars]{Star formation in high-redshift quasars: excess [O{\sc{ii}}] emission in the radio-loud population}
\author[Kalfountzou et al.]
{E. Kalfountzou$^{1}$\thanks{Email: e.kalfountzou@herts.ac.uk},
M. J. Jarvis$^{1,2}$,
D. G. Bonfield$^{1}$ and
M. J. Hardcastle$^{1}$\\
\footnotesize
$^{1}$Centre for Astrophysics, Science \& Technology Research Institute, University of Hertfordshire, Hatfield, Herts, AL10 9AB, UK\\
$^{2}$Physics Department, University of the Western Cape, Bellville 7535, South Africa}
\begin{document}
\date{Accepted 2012 September 9.  Received 2012 August 14; in original form 2012 February 23}


\maketitle
\def\eq#1{\begin{equation} #1 \end{equation}}
\label{firstpage}

\begin{abstract}

We investigate the [O{\sc ii}] emission line properties of 18,508 quasars at $z<1.6$ drawn from the Sloan Digital Sky Survey (SDSS) quasar sample. The quasar sample has been separated into 1,692 radio-loud and 16,816 radio-quiet quasars (RLQs and RQQs hereafter) matched in both redshift and $i'$-band absolute magnitude. 

We use the [O{\sc ii}]$\lambda3726+3729$ line as an indicator of star formation. Based on these measurements we find evidence that star-formation activity is higher in the RLQ population. The mean equivalent widths (EW) for [O{\sc{ii}}] are ${\rm EW}($[O{\sc{ii}}]$)_{\rm RL}=7.80\pm0.30$~\AA\, and ${\rm EW}($[O{\sc{ii}}]$)_{\rm RQ}=4.77\pm0.06$~\AA\, for the RLQ and RQQ samples respectively. The mean [O{\sc{ii}}] luminosities are $\log_{10}[L($[O{\sc{ii}}]$)_{\rm RL}/{\rm W}] = 34.31\pm0.01$ and $\log_{10}[L($[O{\sc{ii}}]$)_{\rm RQ}/{\rm W}]= 34.192\pm 0.004$ for the samples of RLQs and RQQs respectively. Finally, to overcome possible biases in the EW measurements due to the continuum emission below the [O{\sc{ii}}] line being contaminated by young stars in the host galaxy, we use the ratio of the [O{\sc{ii}}] luminosity to rest-frame $i'$-band luminosity, in this case, we find for the RLQs $\log_{10}[L($[O{\sc{ii}}]$)_{\rm RL}/L_{\rm opt}] = -3.89 \pm 0.01$ and $\log_{10}[L($[O{\sc{ii}}]$)_{\rm RQ}/L_{\rm opt}] = -4.011 \pm 0.004$ for RQQs.
However the results depend upon the optical luminosity of the quasar. RLQs and RQQs with the same high optical luminosity $\log_{10}(L_{\rm opt}/$W$)>38.6$, tend to have the same level of [O{\sc{ii}}] emission. On the other hand, at lower optical luminosities $\log_{10}(L_{\rm opt}/$W$)<38.6$, there is a clear [O{\sc{ii}}] emission excess for the RLQs. As an additional check of our results we use the [O{\sc{iii}}] emission line as a tracer of the bolometric accretion luminosity, instead of the $i'$-band absolute magnitude, and we obtain similar results.

Radio jets appear to be the main reason for the [O{\sc{ii}}] emission excess in the case of RLQs. In contrast, we suggest AGN feedback ensures that the two populations acquire the same [O{\sc{ii}}] emission at higher optical luminosities.

\end{abstract}


\section{Introduction}

\subsection{The link between accretion activity and star-formation activity}

It is now clear that black-hole growth, and associated active galactic nuclei (AGN), and the growth of the host galaxy are connected (e.g. \citealt{Magorrian1998, FerrareseMerritt2000, Gebhardt2000, HaringRix2004}). Furthermore, simulations assume that the triggering of the main phase of AGN activity in gas-rich mergers will be accompanied by a major galaxy-wide starburst (e.g. \citealt{DiMatteo2005, Springel2005, Hopkins2008a, Somerville2008}). 

In order to understand the link between the two processes and assess the possibility of the two occurring concomitantly, it is important to quantify and constrain the star-formation activity in quasar host galaxies. From early studies on the star formation -- quasar connection, it was realized that the evolving luminosity density of quasars (e.g. \citealt{Boyle1998, Richards2005, Croom2009}) and the cosmic star-formation history (e.g. \citealt{Madau1996, HopkinsBeacom2006}) are similar (e.g. \citealt{Franceschini1999, Archibald2002}). Supporting evidence has also reported the presence of dust (\citealt{Archibald2001, Page2001, Reuland2004, Stevens2005}), cold gas (e.g. \citealt{Evans2001, Scoville2003, Walter2004, Emonts2011}) and young stars (e.g. \citealt{Tadhunter2005, Baldi2008, Herbert2010}) in powerful radio-loud AGN, which does not reconcile with the picture of low star-forming activity in the ellipticals that usually host these objects (e.g. \citealp{Dunlop2003}). 

Many studies have also attempted to determine the star-formation activity in quasar host galaxies using optical colours (e.g. \citealp{Sanchez2004}) or spectroscopy (e.g. \citealt{Trichas2010, Kalfountzou2011,Trichas2012}). The star formation and quasar connection has also been investigated in recent work (e.g. \citealp{Trichas2009}; \citealp{Serjeant2010}; \citealp{Bonfield2011}) using far-infrared and submillimiter data from the {\em Herschel} Astrophysical Terahertz Large Area Survey (H-ATLAS; \citealp{Eales2010}) and by \citet{Hatziminaoglou2010} using the {\em Herschel} Multi-Tiered Extragalactic Survey (HerMES; \citealp{Oliver2010}), showing that the far-infrared emission is weakly correlated with  the accretion rate for optically selected quasars. In complementary work, \citet{Hardcastle2010} find that the far-infrared emission in radio galaxies appears consistent with the population of non-active galaxies, at least at low radio luminosities, while \citet{Seymour2011} suggest that the star-formation activity in radio galaxy hosts increases with redshift.

\subsection{The radio-loud -- radio-quiet dichotomy}
The radio-loudness dichotomy constitutes a subject that has attracted considerable attention over the years. RLQs are often defined to be the subset of quasars with a radio-loudness satisfying $R_{i}>10$, where $R_{i}=\frac{L({\rm 5GHz})}{L({\rm 4000\AA})}$ \citep{Kellermann1989} is the ratio of monochromatic luminosities measured at (rest frame) 5~GHz and 4000~\AA. The RQQs must minimally satisfy $R_{i}\leq10$. It has long been claimed that RLQs are only a small fraction $\sim$10 per cent of all quasars (e.g. \citealp{Ivezic2002}), with this fraction possibly varying with both luminosity and redshift \citep{Jiang2007}. 

One of the main difficulties in studies aimed at comparing the properties of the two populations is the low fraction of radio-loud sources, which leads to difficulty in defining statistically meaningful samples of RLQs. With the advent of the SDSS database, coupled with the Faint Images of the Radio Sky at Twenty cm survey (FIRST survey; \citealp{Becker1995}), it became possible to select a large and more complete sample of radio-loud AGN. Recent SDSS-based studies (e.g. \citealt{Ivezic2002, White2007, Labita2008}) confirm the presence of a dichotomy in the radio properties of quasar populations.

While the definitive physical explanation of this dichotomy remains confused, a large number of models have been put forward to explain it. Some of them are based on the idea that RLQs require more massive central black holes than RQQs (e.g. \citealt{Dunlop2003, Jarvis2002}; also \citealp{Shankar2010} find this to be redshift dependent) and it has also been suggested that RLQs host more rapidly spinning black holes than RQQs (e.g. \citealt{BlandfordZnajek1977, Punsly1990, Wilson1995, Sikora2007}; but see also \citealp{Garofalo2010}). Apart from the central black hole, differences between the two populations have also been found in their environment and host galaxies. Recently, \citet{Falder2010} showed that radio-loud AGN appear to be found in denser environments than their radio-quiet counterparts at $z\sim 1$, in contrast with previous studies at lower redshifts (e.g. \citealp{McLure2001}). However the differences are not large and may be partly explained by an enhancement in the radio emission due to the confinement of the radio jet in a dense environment (e.g. \citealp{Barthel1996}).

At lower redshift, the SDSS has also been used to investigate differences in the host galaxies, star formation activity and environments for well-defined samples of radio galaxies \citep[e.g.][]{Kauffmann2003, Kauffmann2004, Best2005, BestHeckman2012}. Of particular relevance to the work presented here, \cite{Kauffmann2008} suggest causal connections between the ability of an AGN to produce radio emission and the environment in which is resides, with radio AGN being found in denser regions. This is in qualitative agreement with work on higher redshift quasars discussed above. However, we note that the sample radio galaxies analysed by \cite{Kauffmann2008} generally lie at much lower radio luminosities than the quasars under investigation in this paper, and as noted by \cite{Kauffmann2008} the radio luminosity appears to be loosely correlated with the accretion rate and as such it is difficult to quantitatively compare this work with the studies of higher redshift, higher luminosity, quasars.

If the radio-loudness is due to the physics of the central engine and how it is fueled, and the environment plays a relatively minor role, the quasar properties may be connected with the star formation in their host galaxies \citep[e.g.][]{Herbert2010}. On the one hand, AGN feedback could be stronger in the case of the RLQs due to their higher black hole masses and therefore potentially higher radiation field, reducing the star-formation rate compared to RQQs while on the other hand radio jets could increase the star-formation activity compressing the intergalactic medium \citep[e.g.][]{Croft2006, SilkNusser10}.

In this paper, we present a sample of 18,508 quasars, spectroscopically confirmed by the SDSS and with radio data drawn from the FIRST survey. We use the quasar sample to investigate any possible differences in the star-formation activity of RLQs and RQQs using [O{\sc{ii}}] emission as a star formation tracer. In Section 2 we describe our sample, our selection criteria and our measurements. Section 3 presents our results and in Section 4 we discuss plausible explanations. Finally, in Section 5 we provide a summary of our main results. Throughout this paper, we adopt the following values for the cosmological parameters: H$_{0}=70 {~\rm km~ s^{-1}~Mpc^{-1}}$, $\Omega_{\rm M}=0.3$ and $\Omega_{\lambda}=0.7$.

\section{The sample}

Our quasar sample is drawn from the SDSS Quasar Catalogue seventh data release (DR7) \citep{Schneider2010}. It contains a total of 105,783 spectroscopically confirmed quasars brighter than $ M_{i}=-22.0$ with at least one emission line with full-width half maximum (FWHM) greater than $1000 {\rm~ km~ s^{-1}}$ or with complex absorption features. The full range of redshifts is $z=0.065$ to $z=5.46$, with a mean value of $z\sim 1.49$. For the aims of this paper, where we investigate whether there are any differences in the [O{\sc ii}]$\lambda 3726,3729$\AA\, emission line doublet between radio-loud and radio-quiet quasars, we restrict the redshift range to $z< 1.6$ and impose a signal-to-noise ratio in the equivalent width of the [O{\sc{ii}}] emission line which must be greater than 3.

In order to measure the radio emission we cross match the SDSS quasars with the Faint Images of the Radio Sky at Twenty cm survey (FIRST; \citealp{Becker1995}). Sources are extracted using an elliptical Gaussian fitting procedure \citep{White1997} with a $5\sigma$ detection limit of $\sim {\rm 1~mJy}$. 
To match the quasars to the FIRST radio catalogue, we used a $5''$ arcsec radius from the optical counterpart. However, as quasars often show multiple radio components we created a second subset found using a $30''$ match radius where the sub-structured sources were collapsed into single objects having radio fluxes equal to the sum of the flux of the various components and the position of the final single-source being specified by SDSS. With this method we found 9,133 quasars with matching radius less than $5''$ and 321 extended quasars. However, limiting the sample to only the radio detected sources, we ignore the main bulk of the population. Therefore, to study the entire quasar population in the radio, we also stacked the radio images and detected the median flux densities of undetected sources, following the method of \citet{White2007}.

We separated the quasar sample into RLQs and RQQs based on the radio-loudness, $R_{i}$, parameter. The traditional radio-loud/quiet division is taken to be the radio-to-optical flux ratio of 10. In this work, following \citet{Ivezic2002}, we compute the radio-to-optical flux ratio, $R_{i}$, using the radio flux density at 1.4~GHz and the $i'$-band ($\lambda_{i} = 7480$~\AA) flux density from the SDSS.

In order to check that the radio maps from the FIRST survey do not miss a significant fraction of extended emission around the quasars in our sample due to the relative lack of short baselines of the VLA in B-array, we also estimate the $R_{i}$ parameter using NRAO VLA Sky Survey (NVSS; \citealp{Condon1998}). We find no significant difference in the $R_{\rm 1.4GHz}$ between FIRST and NVSS, with the distribution remaining statistically unchanged. This suggests that the radio emission from the quasars is generally fairly compact.

We must also ensure that the two subsets of RLQs and RQQs have the same optical luminosity and redshift distribution, so that any differences found in the [O{\sc{ii}}] emission-line properties for the RLQ and RQQ samples are not due to differences in their optical properties. As redder passbands measure flux from the part of the spectrum relatively insensitive to recent star formation and also suffer less dust extinction, we use the $i'$-band magnitude to estimate the optical luminosity. We therefore randomly remove RQQs from our parent sample in order to force the same distribution in redshift and $i'$-band luminosity for both the RQQ and RLQ subsamples. It is well known that the continuum luminosity in quasars is a good tracer of the accretion luminosity, as the emission is coming directly from the accretion disk \citep[e.g.][]{Rees1984}. As a result, our matched samples of RLQs and RQQs in optical luminosity and redshift should be also matched in accretion luminosity.

\begin{figure}
\centerline{\hspace{-2.7cm}\includegraphics[scale=0.55]{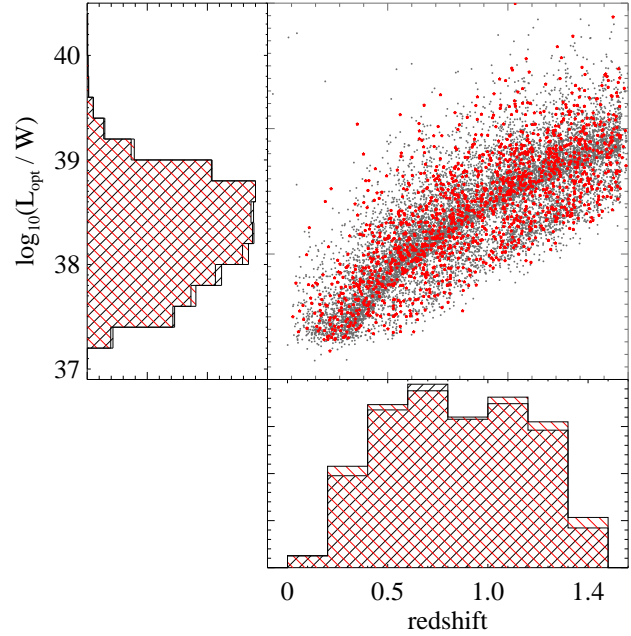}}
\vspace{-2.0cm}
\caption{Distribution of RLQs (red) and RQQs (grey) in luminosity and redshift space. The left and bottom panels show the normalized optical luminosity and redshift histograms.}
\label{fig:Mivsz}
\end{figure}

Figure~\ref{fig:Mivsz} shows the final distribution of the 1,692 RLQs and 16,816 RQQs in the redshift -- optical luminosity ($L_{\rm opt}-z$) plane and their respective histograms. A two-dimensional Kolmogorov-Smirnov (K-S) test applied to the $L_{\rm opt}-z$ distribution shown in Figure~\ref{fig:Mivsz} shows that the distribution of the RLQs and RQQs samples are indistinguishable with a probability that they are drawn from the same underlying distribution $p=0.99$. An one-dimensional K-S test applied to the $\log_{10}L_{\rm opt}$ distribution gives: $D=0.01$ and $p=0.99$ while the one-dimensional K-S test to the redshift distribution gives: $D=0.29$ and $p=0.97$. The mean redshift for the two populations is $z=0.83$ for the RLQs and $z=0.84$ for the RQQs, while the mean luminosity $\log_{10}(L_{\rm opt}/{\rm W}) = 38.30$ for both.

\subsection{Spectral measurements}

In addition to the [O{\sc{ii}}] measurements, which are used to investigate any possible difference in the star formation in RLQs and RQQs, we are also interested in the H$\beta$ and Mg{\sc ii} emission lines and the continuum luminosity at 3000 and 5100~\AA\, because they have been calibrated as virial black hole mass estimators \citep[e.g.][]{McLure2002} and the [O{\sc{iii}}] emission line which has been used  to trace the accretion luminosity in previous studies \citep[e.g.][]{Simpson2005}. 

\begin{figure*}
\centering
\begin{tabular}{c c c}
\includegraphics[scale=0.35]{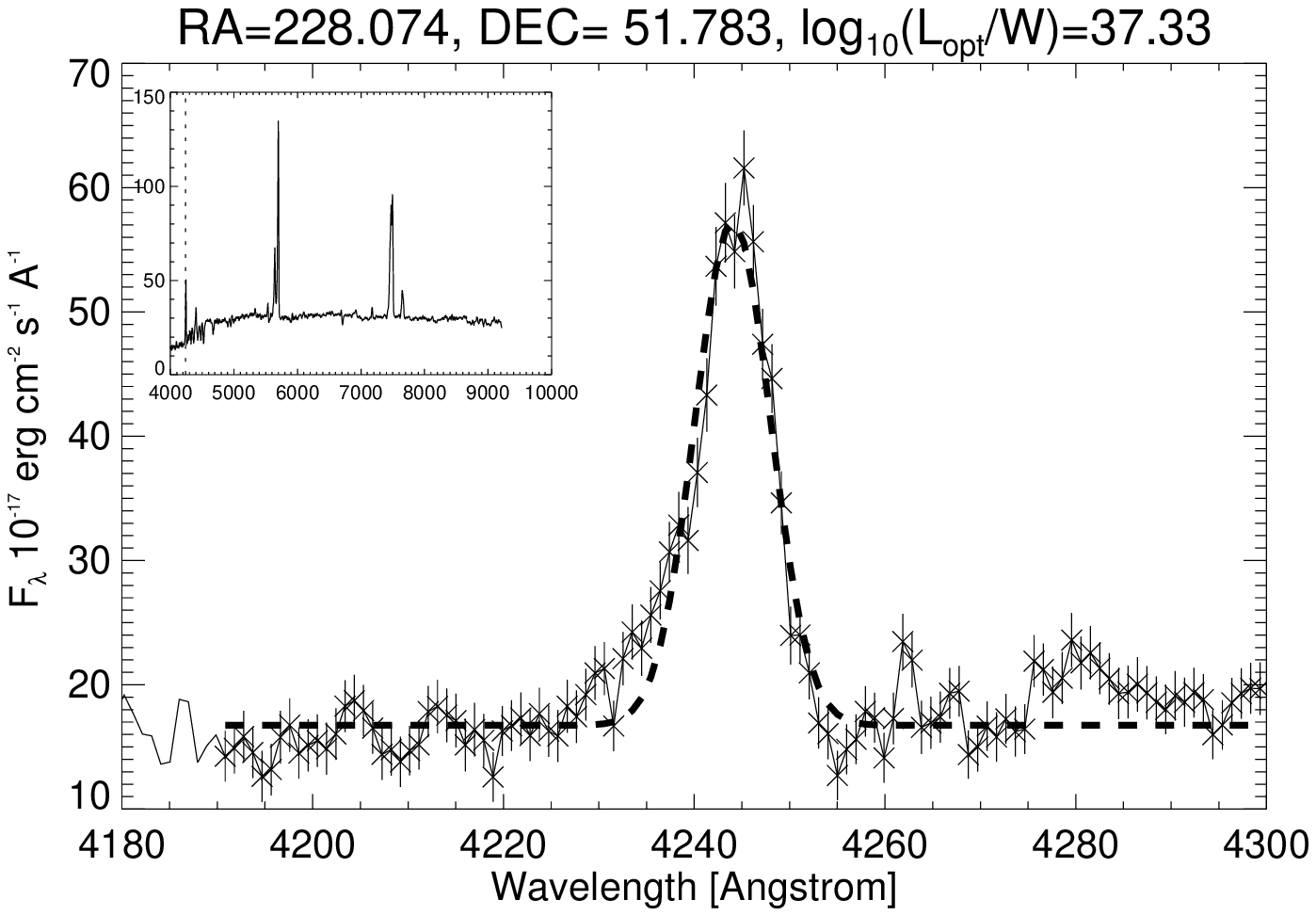} &
\includegraphics[scale=0.35]{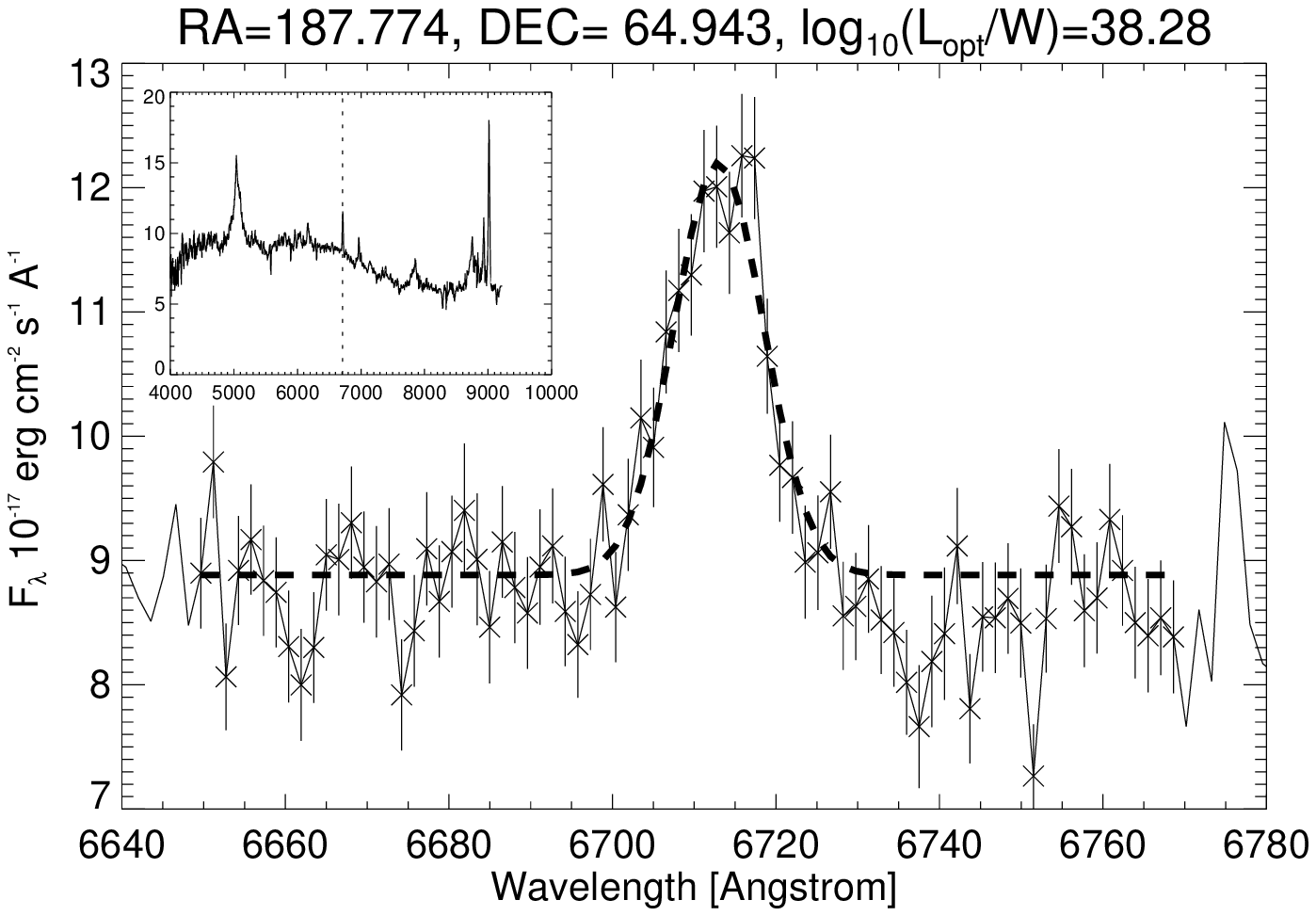} &
\includegraphics[scale=0.35]{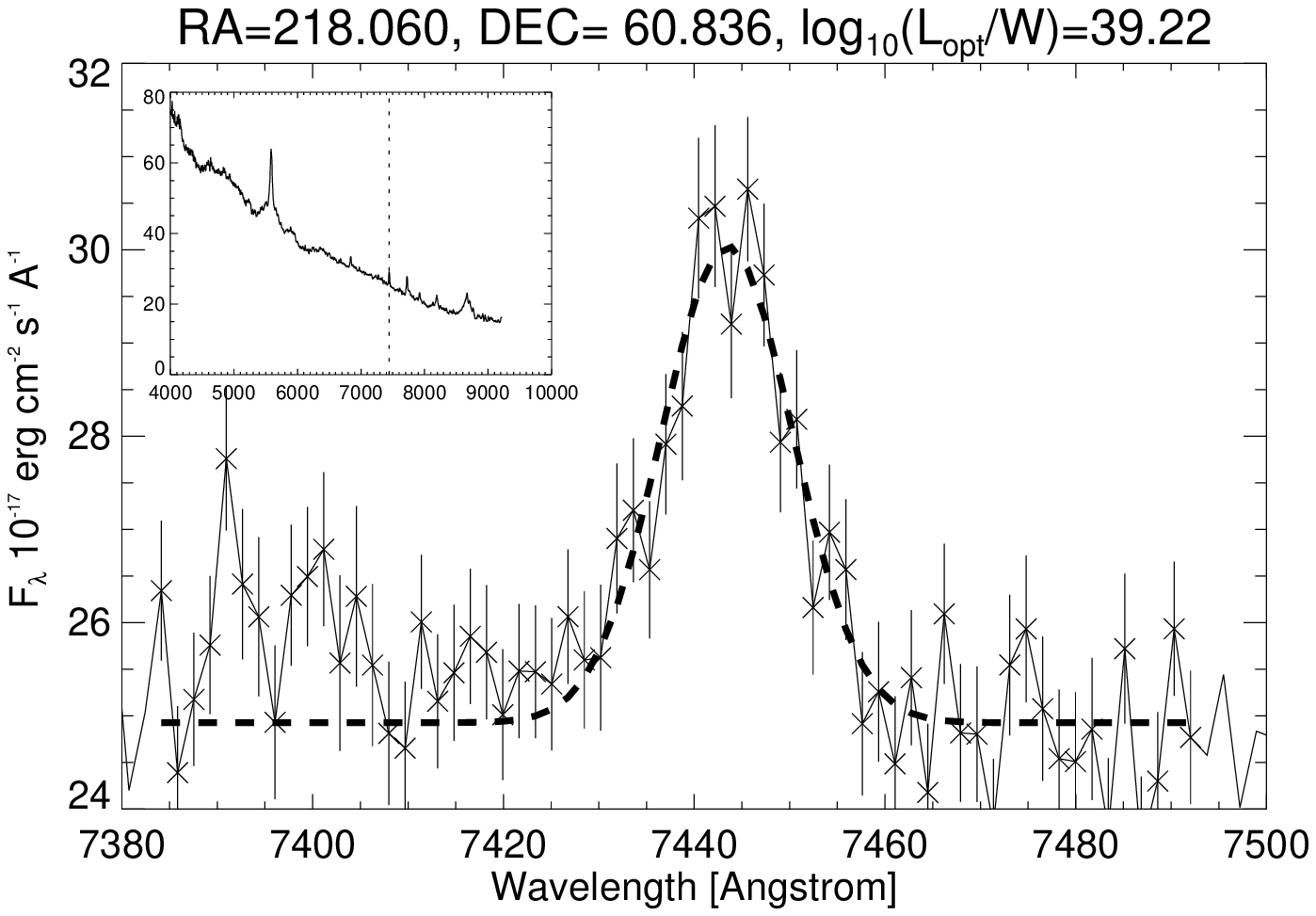}  \\
\includegraphics[scale=0.35]{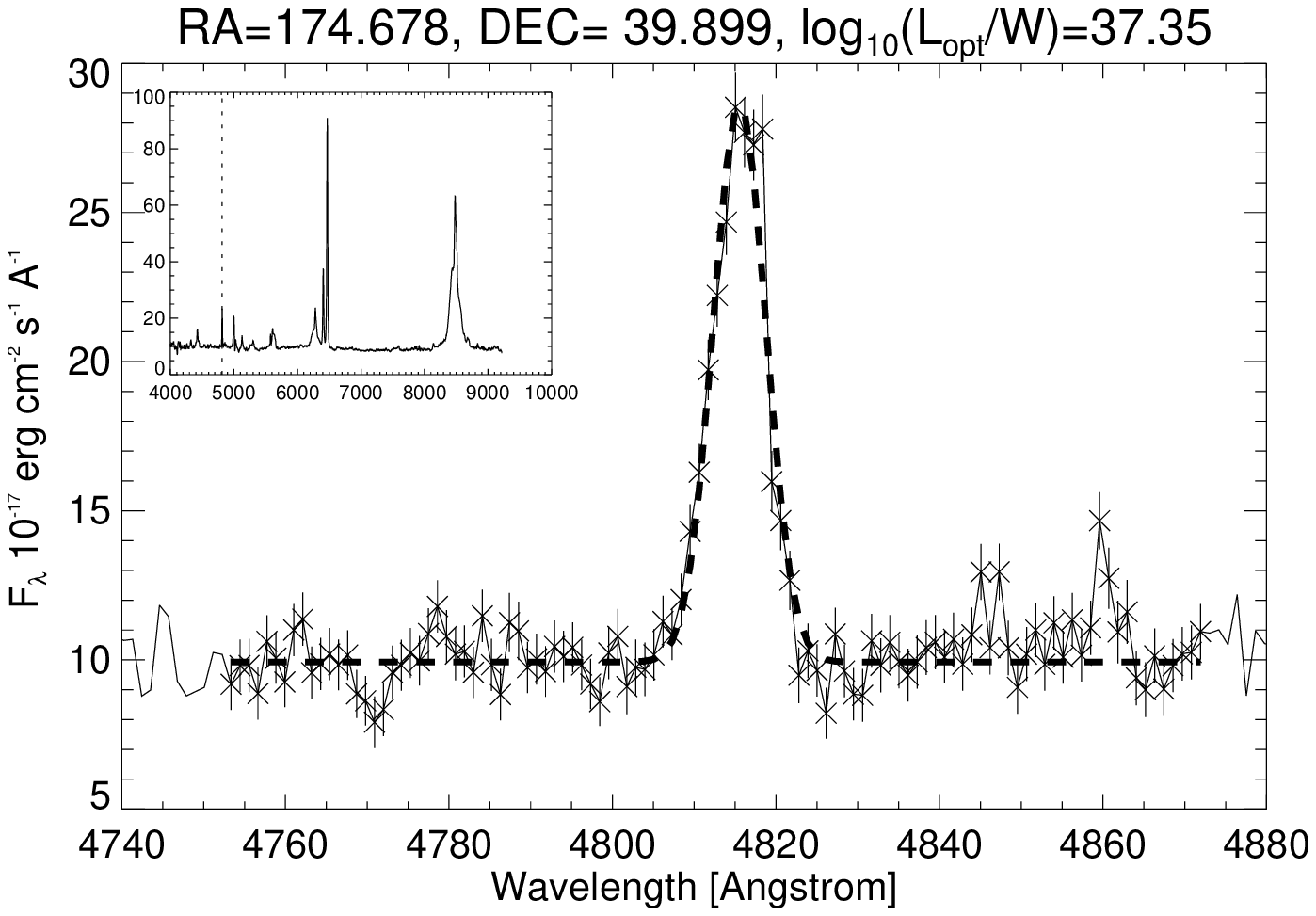} &
\includegraphics[scale=0.35]{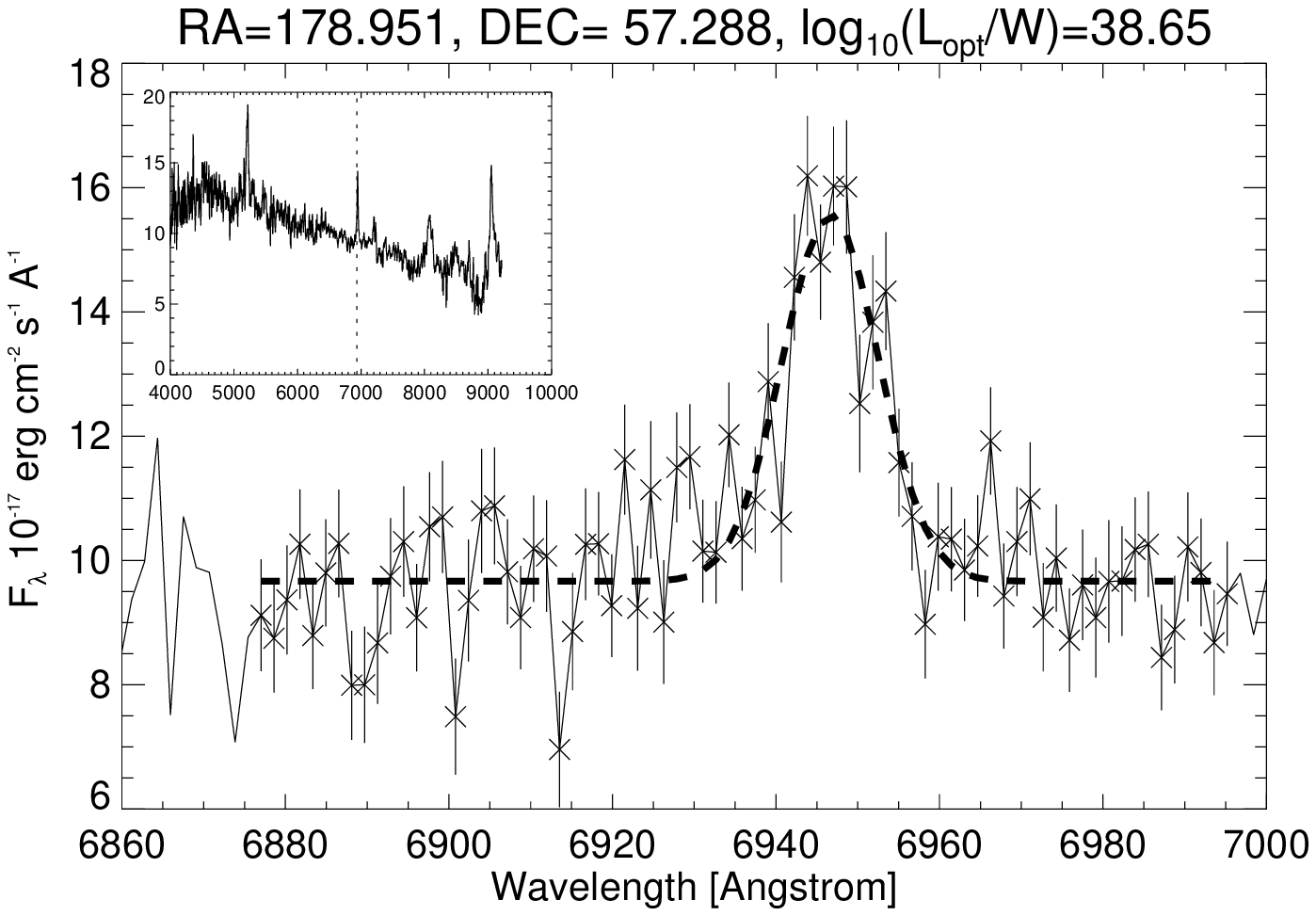} &
\includegraphics[scale=0.35]{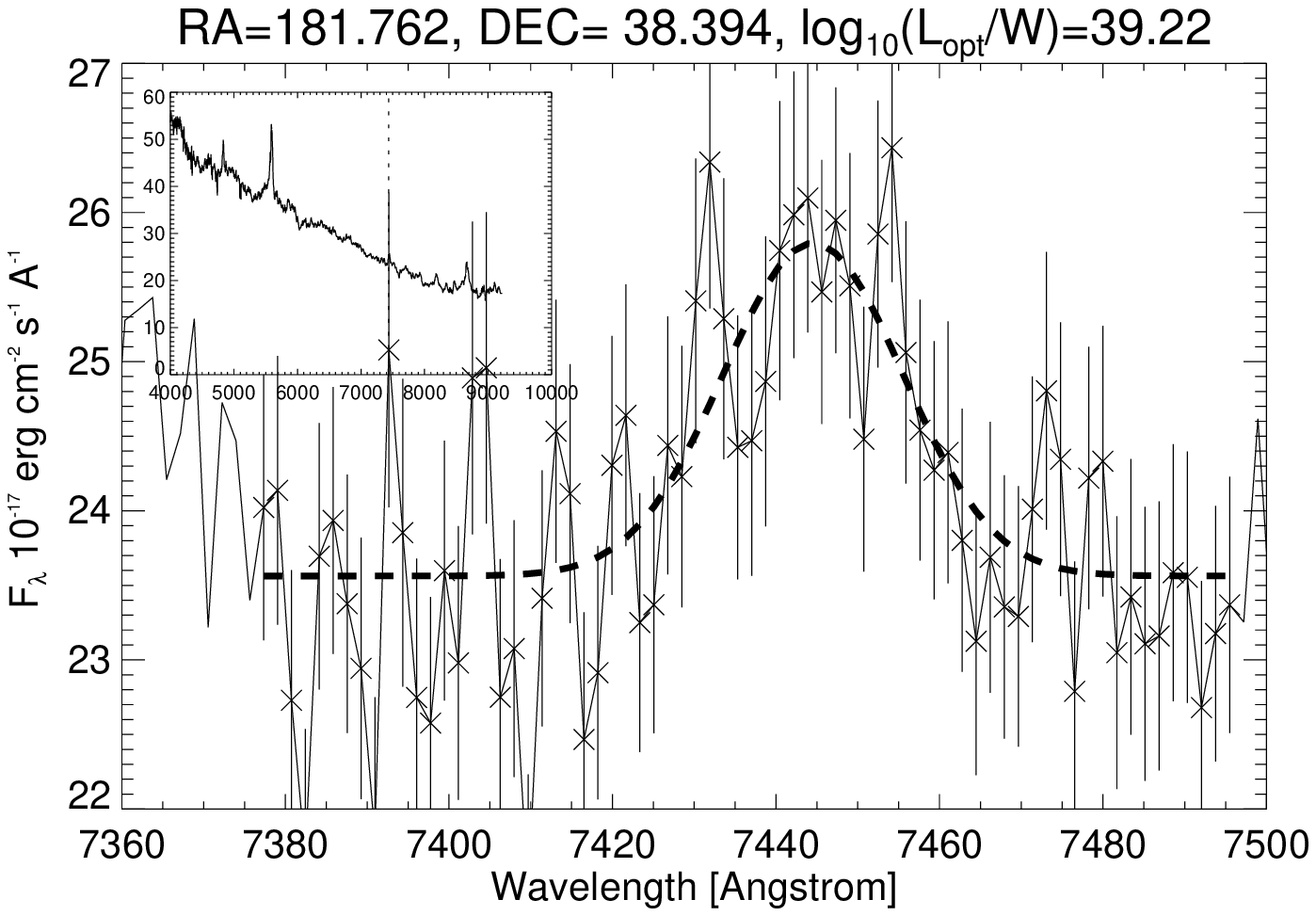} \\
\end{tabular}
\caption{Example spectra (insets) drawn from our samples of RLQs (upper panels) and RQQs (lower panels) with their [O{\sc{ii}}] emission line (main panels). The optical luminosity of the objects is shown above each plot with the luminosity increasing from left to right. The vertical line in the inset identifies the [O{\sc{ii}}] emission line. The dashed black line in the main panels shows the fitted continuum level and Gaussian fit to the emission line. The noise characteristics are also shown for completeness.}
\label{fig:examples}
\end{figure*}

Based on the FWHM measurements we separate the emission lines into narrow (${\rm FWHM}<1000~ {\rm km~ s^{-1}}$) and broad (${\rm FWHM} \geq 1000~ {\rm km~ s^{-1}}$) lines. We fit the narrow emission lines with a single Gaussian while broad lines are fitted with two/three Gaussians. The noise in the spectra is measured using emission line free regions, which cover up to 20 \AA, around each of the emission lines we are interested in. The continuum level was removed by fitting power-law continua $f_{v}\propto v^{\alpha}$, where $\alpha={\rm const}$. However, even in the optical-UV range (1200 - 10000 \AA), the continuum spectrum can not always be represented by a single power law. For this reason we estimate the local power-law shape of each quasar in four ranges, namely, the rest-frame intervals, 1400 - 2200, 2150 - 3200, 2950 - 4300 and 3900 - 5500 \AA\ \citep{Natali1998}. Figure~\ref{fig:examples} shows some examples of the spectra and the [O{\sc{ii}}] emission line fits for RLQs and RQQs with different optical luminosities.

As a check of our automated method, we compare the results of our automatic fitting code with manual measurements. To perform the test we chose a subsample of 100 sources with available Mg{\sc ii}, H$\beta$ and [O{\sc{ii}}] emission line and find that the equivalent widths are indistinguishable with a root-mean square scatter of 0.18, 0.09 and 0.18 around the linear 1-1 relation for the Mg{\sc ii}, H$\beta$ and [O{\sc{ii}}] emission lines respectively.

\section{Results}\label{sec:results}

In this section we investigate whether there are any detectable differences in the [O{\sc{ii}}] emission line properties of the RLQs and the matched sample of RQQs, using the distribution of the [O{\sc{ii}}] equivalent width, the [O{\sc{ii}}] luminosity and the ratio of the [O{\sc ii}] luminosity to the accretion luminosity ($L($[O{\sc{ii}}]$)/L_{\rm opt}$).

\subsection{The [O{\sc{ii}}] emission-line distribution}

\begin{figure}
\includegraphics[scale=0.45]{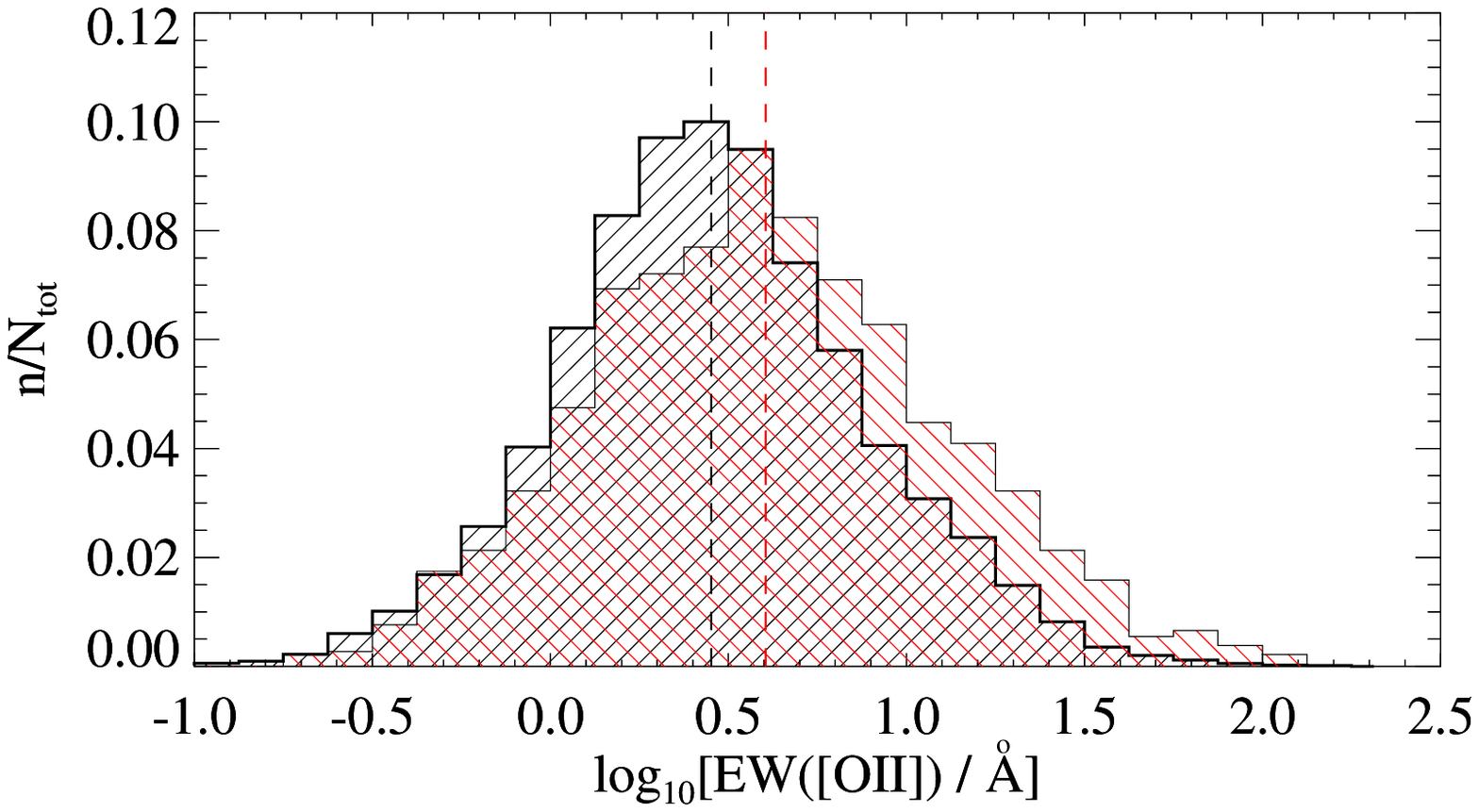}
\includegraphics[scale=0.45]{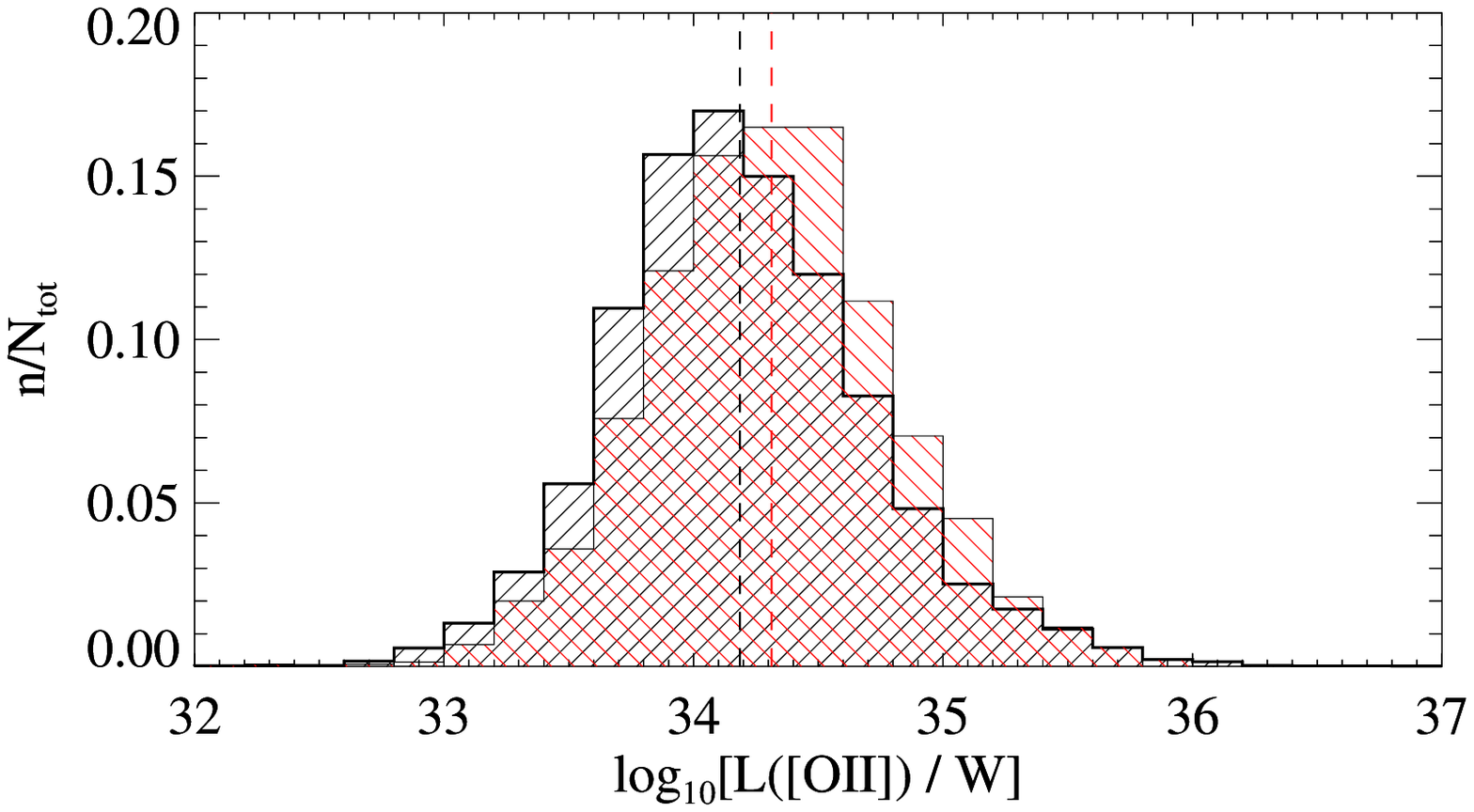}
\includegraphics[scale=0.45]{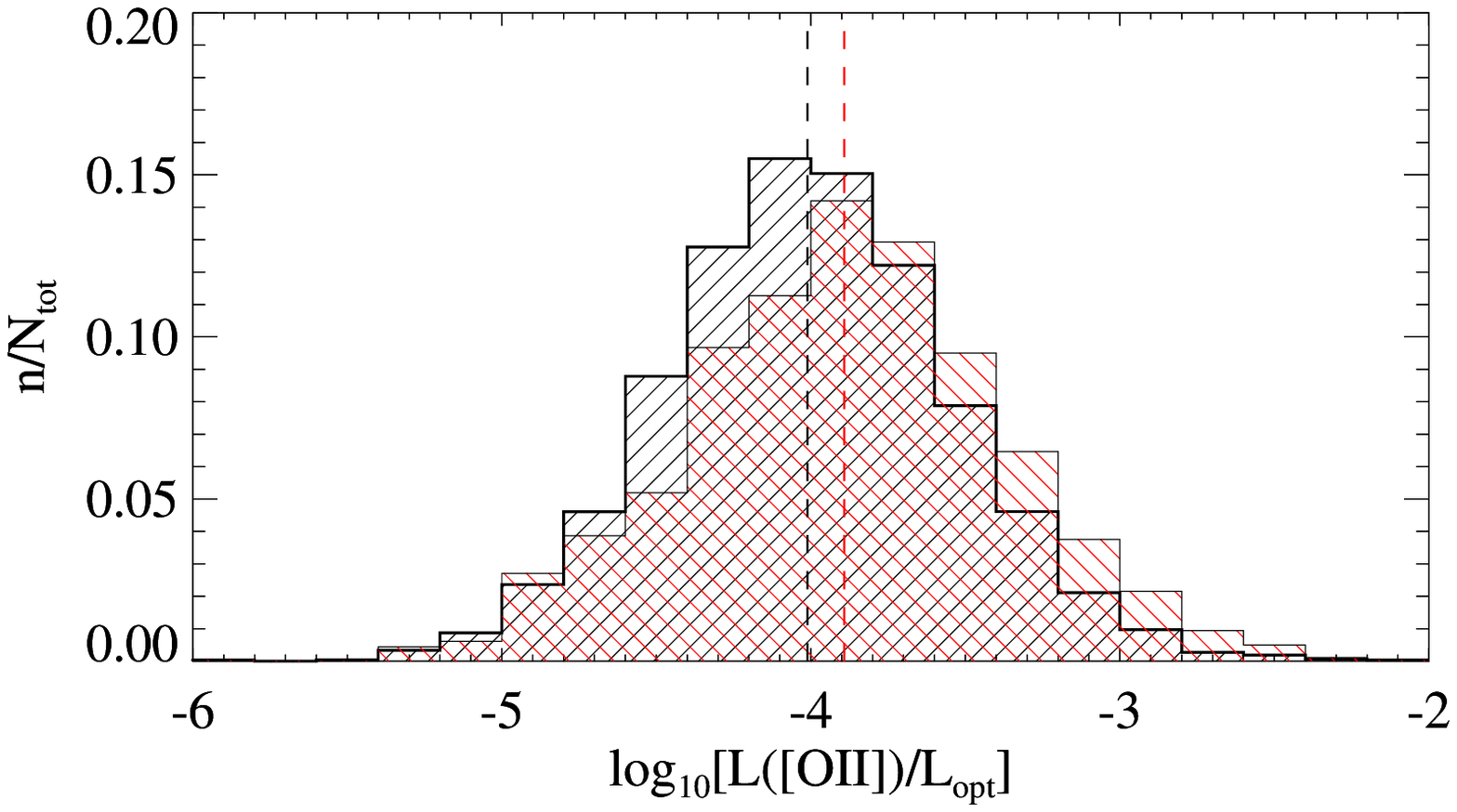}
\caption{Comparison of the $\log_{10}[{\rm EW}($[O{\sc ii}]$)]$, $\log_{10}[L($[O{\sc ii}]$)]$ and $\log_{10}[L($[O{\sc ii}]$)/L_{\rm opt}]$ distribution for the sample of RLQs (red) and the matched sample of RQQs (black). The dashed lines show the mean values for each population.}
\label{fig:EW}
\end{figure}

Figure~\ref{fig:EW} presents a comparison of the [O{\sc{ii}}] emission line properties of the RLQs sample (red histograms) and the matched sample of RQQs (black histograms). The [O{\sc{ii}}] emission line properties that are compared include the distribution of [O{\sc ii}] equivalent width, [O{\sc ii}] luminosity and [O{\sc{ii}}]-optical luminosity ratio ($L($[O{\sc ii}]$)/L_{\rm opt}$), where $L_{\rm opt}$ is determined using the $i'$-band.

The mean equivalent widths are EW([O{\sc ii}])$_{\rm RL}=7.80\pm0.30$~\AA\ and EW([O{\sc ii}])$_{\rm RQ}=4.77\pm0.06$~\AA\ for the RLQ and RQQ samples respectively. The mean [O{\sc ii}] luminosities are $\log_{10}[L($[O{\sc ii}]$)_{\rm RL}/ {\rm W}]=34.31\pm0.01$ and $\log_{10}[L($[O{\sc ii}]$)_{\rm RQ}/ {\rm W}]=34.192\pm0.004$ for the samples of RLQs and RQQs respectively. Finally, in the case of the $L($[O{\sc ii}]$)/L_{\rm opt}$ ratio, we find for the RLQs $\log_{10}[L($[O{\sc ii}]$)/L_{\rm opt}]_{\rm RL}=-3.89\pm0.01$ and $\log_{10}[L($[O{\sc ii}]$)/L_{\rm opt}]_{\rm RQ}=-4.011\pm0.004$ for RQQs. 
We therefore find that the RLQs have different mean values for RQQs in all of the parameters. Recall that the RLQ and RQQ samples are matched in $L_{\rm opt}$, so these results imply that RLQs have higher [O{\sc ii}] emission for a fixed value of $L_{\rm opt}$. We have applied an one-dimensional K-S test for each measurement related to [O{\sc ii}] for the RLQs and RQQs. The K-S test returns a probability of $p=1.12\times 10^{-27}$, $p=3.24\times 10^{-22}$ and $p=8.77\times 10^{-23}$ and a test statistic of $D=0.15$, $D=0.13$ and $D=0.14$ for EW([O{\sc ii}]), $\log_{10}[L($[O{\sc ii}]$)]$ and $\log_{10}[L($[O{\sc ii}]$)/L_{\rm opt}]$ respectively; thus we can reject the null hypothesis that the various measurements related to the [O{\sc ii}] emission are drawn from the same underlying distributions. Based on the distribution histograms, the RLQs have higher [O{\sc{ii}}] emission than that of their RQQ counterparts.

\subsection{The relation between the [O{\sc{ii}}] emission line and $L_{\rm 1.4GHz}$}

Previous studies have found a connection between the narrow emission line luminosity and the radio luminosity in radio galaxies (e.g. \citealt{Rawlings1989, RawlingsSaunders1991, McCarthy1993, Willott1999, Jarvis2001}). The general assumption here is that the narrow line region (NLR) is predominantly photoionised by the quasar and the narrow line luminosity is telling us about the intrinsic quasar luminosity. 

\begin{figure}
\hspace{-0.5cm}
\includegraphics[scale=0.5]{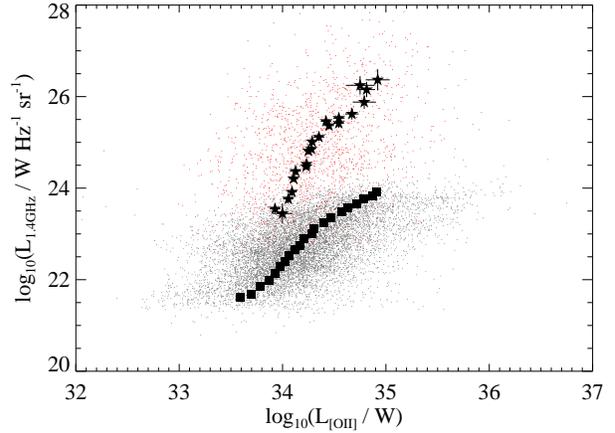}
\caption{Luminosity at 1.4~GHz ($L_{\rm 1.4GHz}$) versus luminosity of [O{\sc{ii}}] $[L($[O{\sc ii}]$)]$ for the RLQ and RQQ subsamples. Red dots represent the RLQs and grey the RQQs. Black stars show the mean values for each optical bin for RLQs while black squares do the same for RQQs.}
\label{fig:L14}
\end{figure}

In Figure~\ref{fig:L14} we show the [O{\sc{ii}}] emission line luminosity against 1.4~GHz radio luminosity. The strong correlation between narrow emission line luminosity and radio luminosity is clearly seen. The trend fit to the whole quasar sample using ordinary least squares (OLS) bisector \citep{Isobe1990} gives a correlation of $\log_{10}[L($[O{\sc ii}]$)/{\rm W}]=(18.94\pm0.15)+(0.67\pm0.01) \log_{10}(L_{\rm 1.4GHz}/ {\rm W~Hz}^{-1}~{\rm sr}^{-1})$. Splitting the sample into the RLQ and RQQ sub-samples the equations for the two populations are:
\\
\\
$\log_{10}[L($[O{\sc ii}]$)]=(19.20\pm 0.47)+(0.61\pm0.02) \log_{10}(L_{\rm 1.4GHz})$\\	
\\
and
\\
\\
$\log_{10}[L($[O{\sc ii}]$)]=(14.88\pm0.14)+(0.85\pm0.01) \log_{10}(L_{\rm 1.4GHz})$	
\\
\\
for RLQs and RQQs respectively. 

The positive correlation  of [O{\sc{ii}}] emission line and radio luminosity is in agreement with previous studies \citep[e.g.][]{Willott1999, Jarvis2001}. \citet{Fernandes2011} performed an analogous study for powerful radio galaxies at 151~MHz and they found a strong positive relation $L($[O{\sc ii}]$) \propto L_{\rm 151MHz}^{0.52\pm0.01}$. The slightly different slope may be explained by the different radio frequencies that are used -- 151~MHz for the \citet{Fernandes2011} study and 1.4~GHz in our case -- or due to the different selected sources -- powerful radio galaxies for the \citet{Fernandes2011} study and quasars in our study. However, it is clear that the relation between radio luminosity and [O{\sc ii}] luminosity is different for the RQQs compared to the RLQs; this again leads to the suggestion that an additional effect is at work for ionizing [O{\sc ii}] at least.

\subsection{The relation between the $L($[O{\sc{ii}}]$)$ emission line and $L_{\rm opt}$}

\begin{figure}
\hspace{-0.5cm}
\includegraphics[scale=0.5]{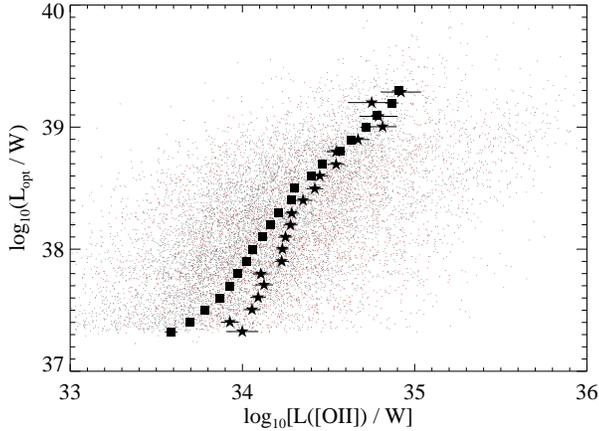}
\caption{Optical luminosity, $L_{\rm opt}$, versus [O{\sc ii}] luminosity, $L($[O{\sc ii}]$)$. Symbols are as in Figure~\ref{fig:L14}.}
\label{fig:LoptvsLOII}
\end{figure}

In Figure~\ref{fig:LoptvsLOII} we show the [O{\sc ii}] emission line luminosity versus the optical luminosity. A strong correlation between the narrow emission line luminosity and the optical luminosity is clearly seen with the OLS bisector fit for the two populations being:
\\
\\
$\log_{10}(L_{\rm opt})=(3.61\pm 0.64)+(1.01\pm 0.02) \log_{10}[L($[O{\sc ii}]$)_{\rm RL}]$	
\\
\\
and 
\\
\\
$\log_{10}(L_{\rm opt})=(5.76\pm 0.23)+(0.95\pm 0.01) \log_{10}[L($[O{\sc ii}]$)_{\rm RQ}]$	
\\
\\
for RLQs and RQQs respectively. It is worth noting that at high optical luminosities [$\log_{10}(L_{\rm opt}/$W$)>38.6$] the RLQs and RQQs show a similar distribution in $\log_{10}[L($[O{\sc ii}]$)]$. However, at lower optical luminosities, the RLQs typically have greater [O{\sc ii}] luminosity than the RQQs.

For the quasars in our sample, the black-hole mass $M_{\rm BH}$\footnote{For quasars up to redshift of 1.6 we have used the \cite{McLure2002} relations based on the FWHM of H$\beta$ and Mg{\sc ii} emission lines to estimate the black hole masses.} is strongly correlated with the optical luminosity. Using the $L_{\rm opt}-M_{\rm BH}$ correlation for the entire sample ($M_{\rm BH}\propto L_{\rm opt}^{1.18\pm0.02}$), $L_{\rm opt}= 10^{38.6}$~W  corresponds to $M_{\rm BH}=10^{8}$~M$_{\odot}$. As suggested by previous studies RLQs tend to have black holes more massive than $10^{8}$~M$_{\odot}$. Above the $10^{38.6}$~W optical luminosity, all the RQQs in our sample have black holes more massive than $10^{8}$~M$_{\odot}$, although RLQs continue to have a higher mean black-hole mass. 

\begin{figure}
\centering
\includegraphics[scale=0.48]{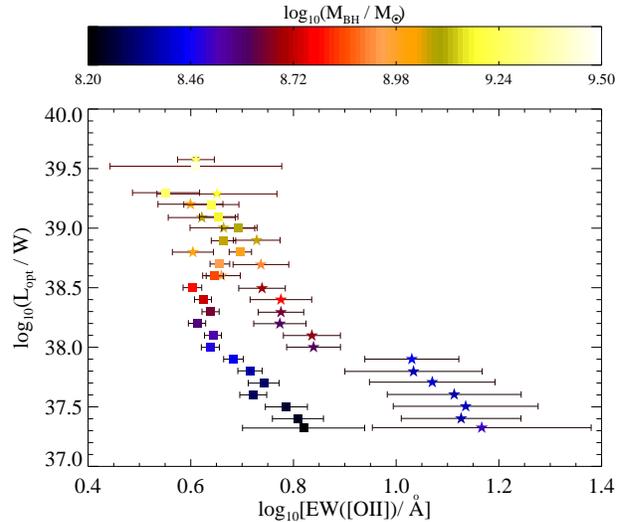}
\caption{Optical luminosity, log$_{10}(L_{\rm opt})$, versus the logarithm of the [O{\sc{ii}}] equivalent width, $\log_{10}[{\rm EW}($[O{\sc{ii}}]$)]$. The colour scale corresponds to the black hole mass estimate. The stars represent the RLQs and the squares the RQQs.}
\label{fig:sumup}
\end{figure}

Figure~\ref{fig:sumup} provides a clearer view of all these parameters where we show the optical luminosity versus [O{\sc{ii}}] equivalent width for the RLQs and RQQs, with black-hole mass indicated using a colour scale. It is obvious that the two samples tend to have the same equivalent width distribution for $\log_{10}(L_{\rm opt} / $~W$)>38.6$, whereas at $\log_{10}(L_{\rm opt} / $~W$)<38.6$ there appears to be a pronounced difference. A possible explanation for these differences between the two populations, and for different optical luminosities (or black hole masses), could be two different mechanisms that are correlated with star formation. At $L_{\rm opt} = 10^{38}$~W, the [O{\sc ii}] emission in the RLQs is roughly 0.4~dex higher than that in the RQQs, corresponding to a star-formation rate of $\sim 4$~M$_{\odot}$~yr$^{-1}$ \citep{Gilbank10}. To explain the [O{\sc{ii}}] emission for optical luminosities lower than $10^{38.6}$~W we need a positive mechanism that increases the star formation in RLQs but not in RQQs. The obvious candidate mechanism is that the jets observed in RLQs but not in RQQs play a critical role. On the other hand, for the higher optical luminosities, we need a mechanism that decreases the [O{\sc{ii}}] emission in RLQs so that the excess with respect to RQQs is destroyed. As the optical limit is correlated with the black hole mass, the explanation we need could be provided by negative AGN feedback. Both of these mechanisms and their connection with [O{\sc{ii}}] emission are discussed in Section 5.

\subsection{[O{\sc{iii}}] as a tracer of accretion luminosity}

In this work we use the $i'$-band magnitude in order to estimate the optical luminosity. We have selected the $i'$-band as an AGN tracer due to the fact that it lies at the redder part of the optical spectrum which is less sensitive to the recent star formation. However, any enhancement in the [O{\sc{ii}}] luminosity needs to be separated from any possible general enhancement in the emission-line luminosity in radio-loud quasars \citep[e.g.][]{Kauffmann2008}. In this section, we therefore test whether our results are robust to another tracer of the accretion luminosity, namely the [O{\sc{iii}}]$\lambda 5007$ emission line. The [O{\sc{iii}}]$\lambda 5007$ line is usually the strongest emission-line in the optical spectra of type-2 AGN, and it is believed to be ionized predominantly by the photons emanating from the accretion disk. The [O{\sc iii}] emission line was also used by \cite{Silverman2009} to statistically remove the component of the detected [O{\sc{ii}}] emission line that can be attributed to the AGN. In order to test the robustness of our results, we use the [O{\sc iii}] luminosity instead of the optical $i'$-band luminosity and conduct a similar analysis. However, we note that \cite{Fine2011} demonstrated that [O{\sc iii}] may not be a robust tracer of the AGN bolometric luminosity and may be affected by orientation. 

\begin{figure}
\begin{tabular}{c}
\hspace{-0.8cm}
\includegraphics[scale=0.5]{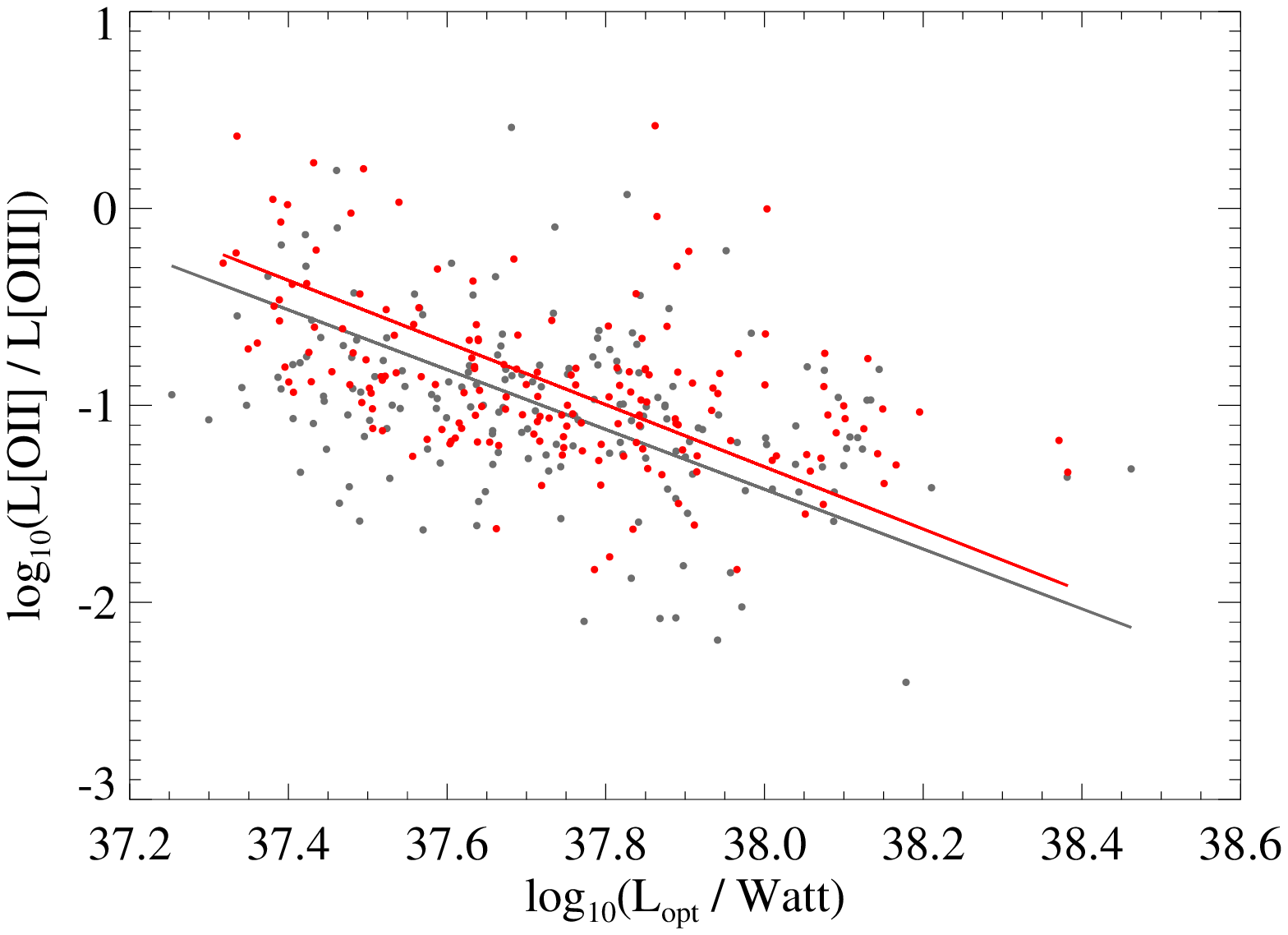} \\
\hspace{-0.8cm}
\includegraphics[scale=0.5]{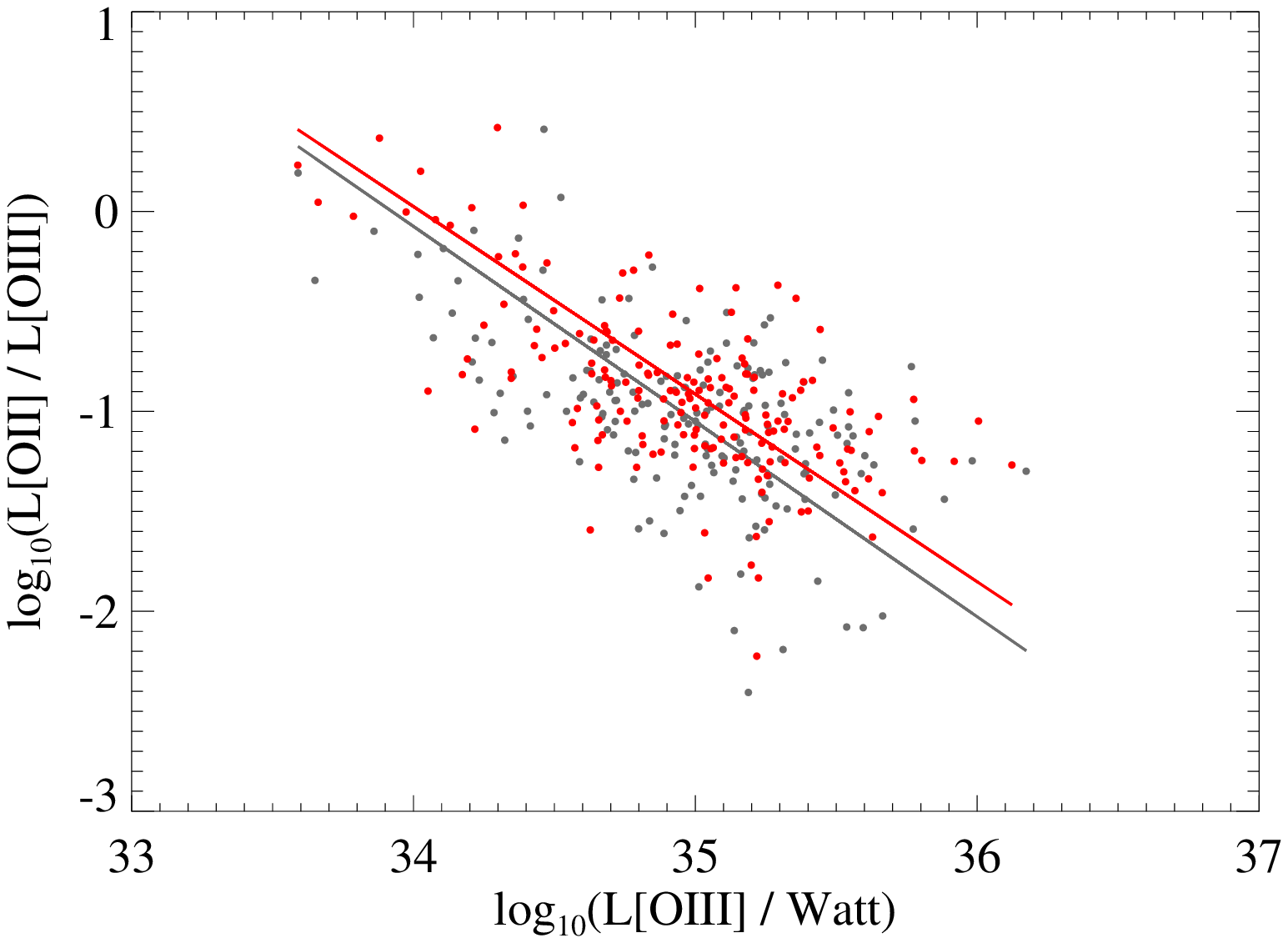} \\
\end{tabular}
\caption{$\log_{10}(L($[O{\sc{ii}}]$)/L($[O{\sc{iii}}]$))$ ratio versus optical luminosity (top) and [O{\sc{iii}}] luminosity (bottom). The red points represent the RLQs and the grey points the RQQs matched in [O{\sc{ii}}] and optical luminosity. The solid lines with the same colours are the OLS bisector fits for the two samples.}
\label{fig:LoptvsOIII}
\end{figure}

In our sample of 18,508 QSOs, 1,629 (1,364 RQQs and 265 RLQs) have the [O{\sc iii}] emission line accessible in the optical spectra. In order to ensure that the [O{\sc{ii}}] emission excess is not biased due to the selection of optical luminosity as an AGN tracer, we follow the same method of the matched subsamples using the [O{\sc iii}] luminosity as a tracer. We separate the sample with available [O{\sc iii}] emission line measurements into RLQs and RQQs with indistinguishable [O{\sc iii}] and optical luminosity distributions. The subsamples contain 237 RLQs and the same number of RQQs. The mean values of the [O{\sc iii}] luminosity for the two populations are $<\log_{10}[L($[O{\sc{iii}}]$)_{\rm RL} / {\rm W}]>=34.957\pm 0.011$ and $<\log_{10}[L($[O{\sc{iii}}]$)_{\rm RQ} / {\rm W}]>=34.958\pm 003$, while the one-dimensional K-S test returns a null probability of $p=0.998$. Figure~\ref{fig:LoptvsOIII} shows the comparison of [O{\sc iii}] and optical luminosity to the $\log_{10}(L($[O{\sc{ii}}]$)/L($[O{\sc{iii}}]$))$ ratio for these RLQ and RQQ subsamples. Assuming that both optical and [O{\sc{iii}}] luminosity are good tracers of AGN power, it is clear that the $\log_{10}(L($[O{\sc{ii}}]$)/L($[O{\sc{iii}}]$))$ ratio for both populations decreases strongly with increasing optical and [O{\sc{iii}}] luminosity, in agreement with \cite{Kim2006}. However, in the case of RLQs the $\log_{10}(L($[O{\sc{ii}}]$)/L($[O{\sc{iii}}]$))$ ratio is higher indicating the same result as found using only the optical luminosity matched samples.

\begin{figure}
\includegraphics[scale=0.45]{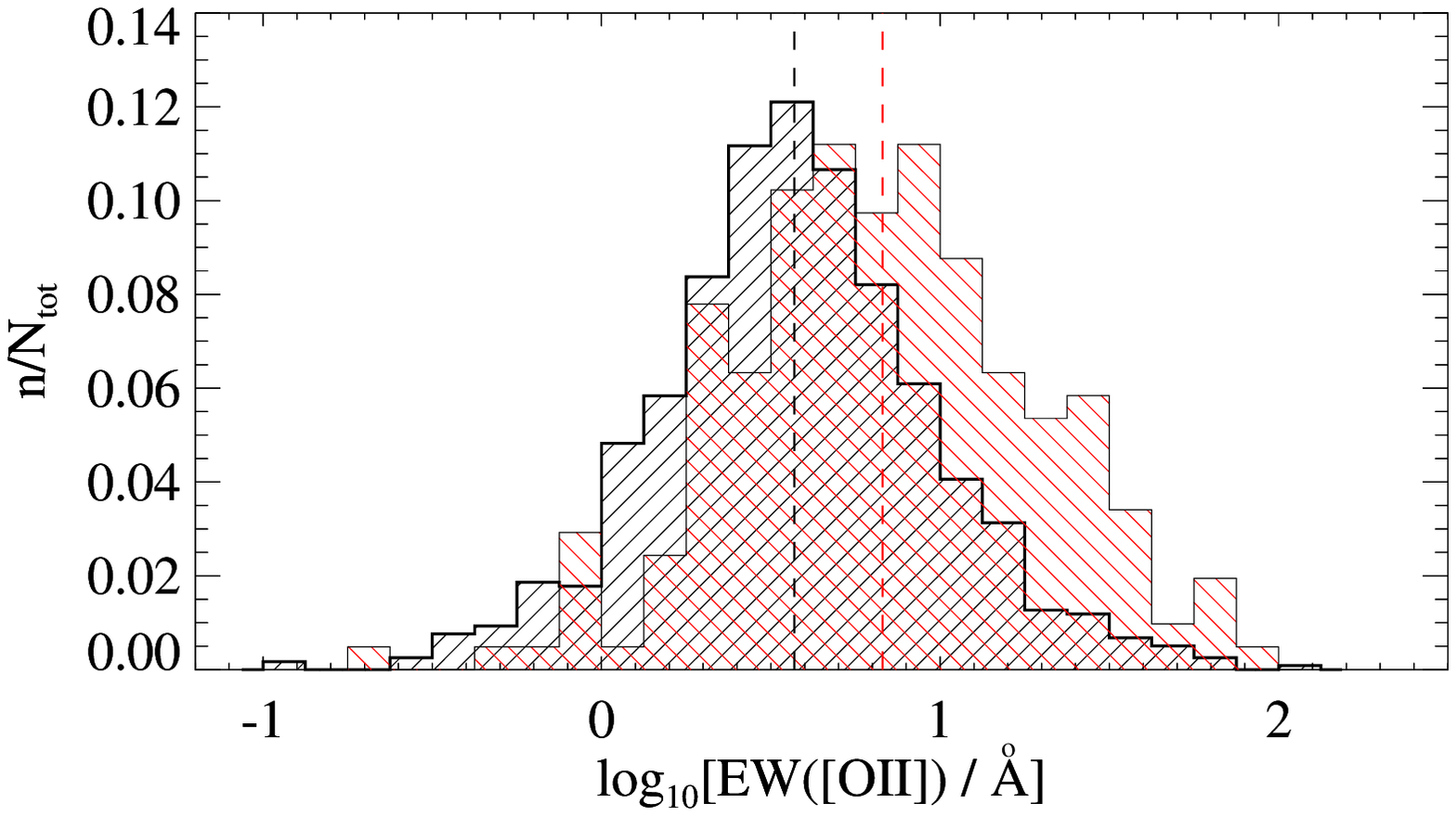}
\includegraphics[scale=0.45]{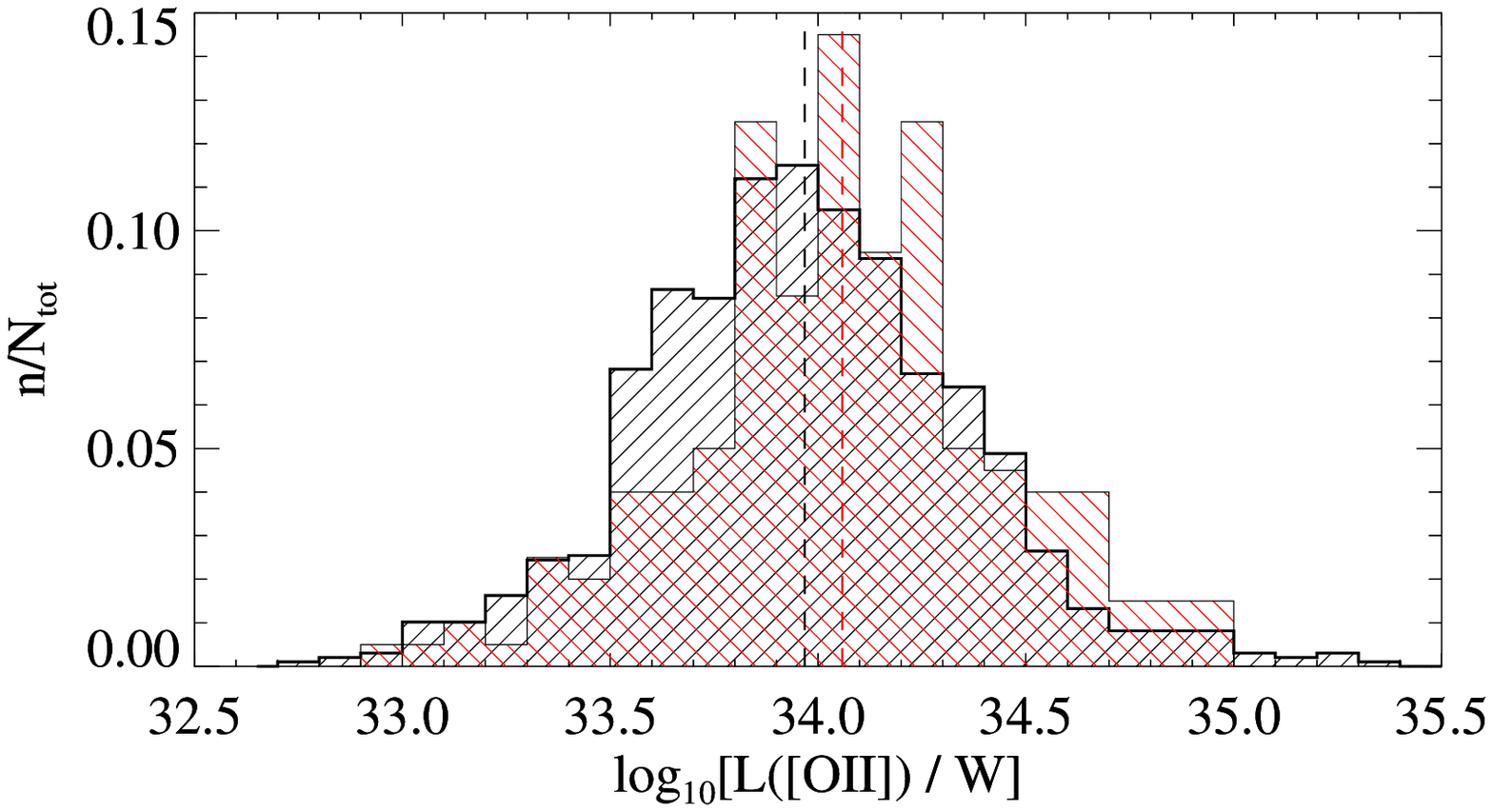}
\includegraphics[scale=0.45]{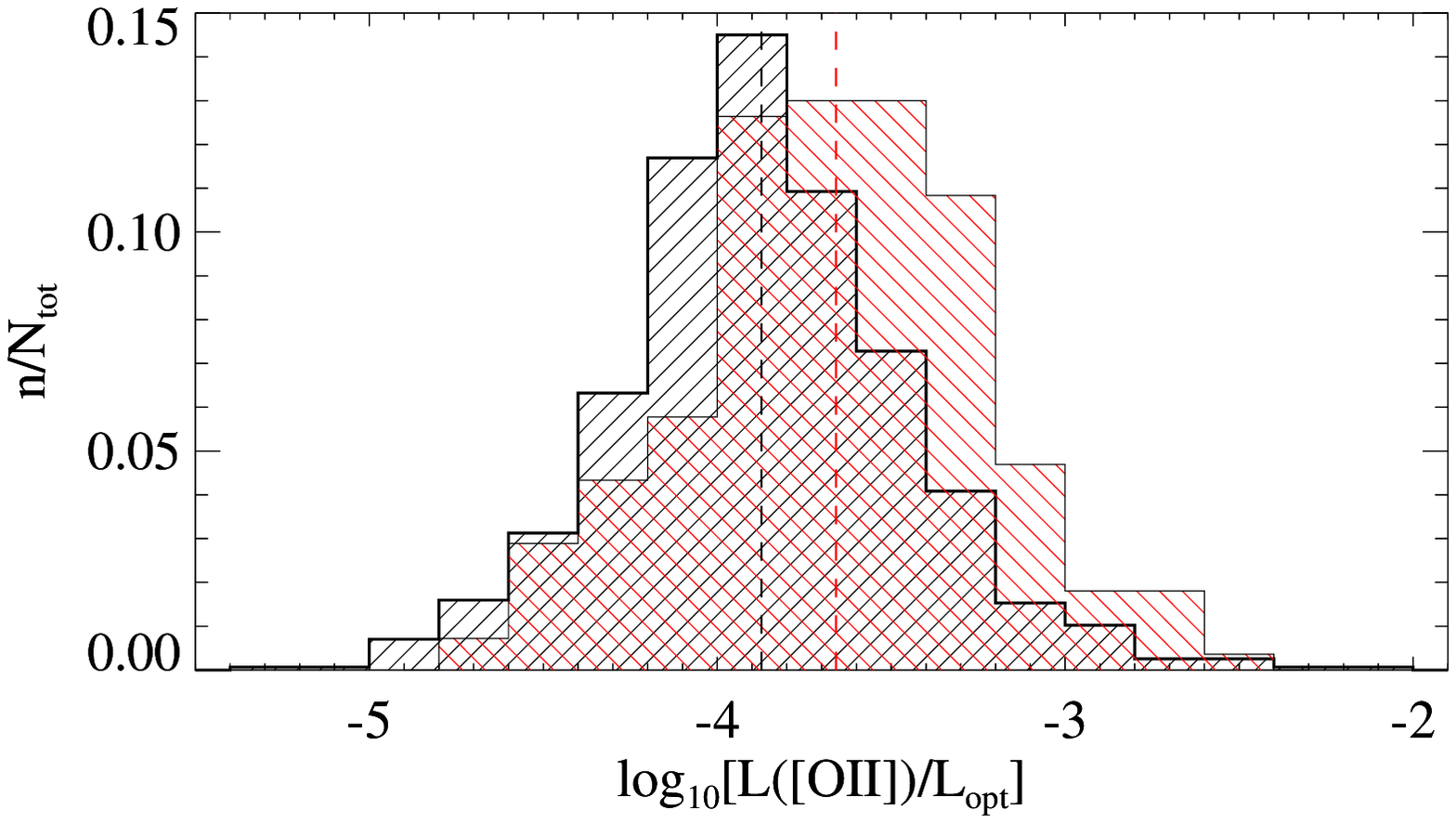}
\caption{Comparison of the $\log_{10}$EW$_{\rm [OII]}$, $\log_{10}[L($[O{\sc{ii}}]$)]$ and $\log_{10}[L($[O{\sc{ii}}]$)/L_{\rm opt}]$ distribution for the samples of RLQs (red) and RQQs (black) with the same [O{\sc iii}] luminosity distribution. The dashed lines show the mean values for each population.}
\label{fig:EW_OIII}
\end{figure}

In Figure~\ref{fig:EW_OIII} we show the comparison of the [O{\sc{ii}}] emission line properties of the RLQs and the matched sample of RQQs. Using the [O{\sc iii}] luminosity as an estimator of AGN luminosity, the [O{\sc{ii}}] emission excess remains for the RLQs. We have applied a one-dimensional Kolmogorov-Smirnov test for each of the histograms in Figure~\ref{fig:EW_OIII} between the RLQs and RQQs. The K-S test returns a probability of $p=9.90\times 10^{-12}$, $p=1.37\times 10^{-6}$ and $p=5.32\times 10^{-10}$ and a test statistic of $D=0.28$, $D=0.15$ and $D=0.26$ for EW$_{\rm [OII]}$, $\log_{10}[L($[O{\sc{ii}}]$)]$ and $\log_{10}[L($[O{\sc{ii}}]$)/L_{\rm opt}]$ respectively, thus the null hypothesis, that the three distributions are drawn from the same underlying distributions, can be ruled out. Therefore, the  matched samples in [O{\sc iii}] luminosity and redshift demonstrate that our results are robust against correlations between radio luminosity and [O{\sc iii}] emission and the corresponding accretion rate derived from these quantities.

\section{Discussion}\label{sec:discussion}

In this section we discuss the mechanisms that may contribute to the observed difference in [O{\sc{ii}}] emission between the RLQ and RQQ populations. 

\subsection{Mergers and triggering of activity}

The early claims that quasars have a high incidence of close companions led many studies to suggest a connection between quasar activity, interactions and mergers \citep[e.g.][]{Urrutia2008, Shankar2010, RamosAlmeida2012}. It has been suggested that major mergers of gas-rich galaxies are an efficient mechanism to feed supermassive black holes (SMBHs). This suggestion is supported by observations of quasar hosts undergoing mergers and interactions \citep[e.g.][]{Kauffmann2003, Urrutia2008}. Although we note that other deep field studies of the hosts of moderate luminosity AGN have found no significant evidence for high merger rates or interactions \citep{Sanchez2004} suggesting that secular processes may trigger these AGN (e.g. bars, slow cold gas accretion, minor interactions, supernova explosions; for reviews, see \citealt{Kormendy2004, Martini2004, Jogee2006}). 
Mergers or inflows of cold gas by other mechanisms not only feed the black hole, but provide a new reservoir of available gas capable of producing a sudden burst of star formation. Based on the scenario of a direct link between mergers and quasars \citep[e.g.][]{Sanders1988}, strong interactions or major mergers between gas-rich galaxies fuel starburst activity and also provide some of the gas for the black hole accretion event. \citet{Bennert2008} suggest that most QSO host galaxies are involved in mergers with accompanying starbursts but usually several hundreds of Myr pass after the merger until the activity is triggered (see also \citealt{Hopkins2006, Wild2010}). This would explain the similarities in [O{\sc ii}] emission for the RLQs and RQQs, at least at high luminosities.

Major galaxy mergers are also expected to lead to black holes with a high value of spin while minor mergers tend to lead to low values of spin. According to the spin paradigm \citep{Wilson1995} the jet production and power are correlated with the angular momentum of a spinning black hole. The spin paradigm has often been used to explain the radio loudness dichotomy, implying that RLQs have black holes that spin more than RQQs.

Based on the above theoretical background, mergers seem to be associated with the AGN activity and the black hole spin which is used to explain the radio loudness dichotomy. However, can mergers on their own explain the [O{\sc{ii}}] emission difference between RLQs and RQQs at low optical luminosity and the similarity in [O{\sc{ii}}] emission at high optical luminosity as found in Section~\ref{sec:results}? 

\citet{Dotti2010} have proposed a model based on gas-rich mergers between nearly equal mass disc galaxies, `wet mergers', that result mostly in radio-quiet AGN residing in elliptical or spiral galaxies with `extra light'. The available gas and the interactions trigger a starburst to create the `extra light' at the center of the galaxy. On the other hand, major gas-poor mergers, `dry mergers', are expected to lead to a radio-loud AGN hosted by an elliptical galaxy. In agreement with this, \cite{Dunlop2003} have shown that radio-loud AGN are predominantly hosted by elliptical galaxies. However, our results do not support this, because if RLQs are the result of major gas-poor merger then the available gas to trigger a starburst is not enough and the estimated [O{\sc{ii}}] emission excess must be the result of a different mechanism. One possibility is that the high spin black hole produced by the merger powers the radio jets which compress the interstellar gas, triggering star-formation activity during the QSO phase.

\subsection{Alignment effect and positive feedback}

One important aspect of AGN is the role that they play in possible triggering of extensive star formation in a multi-phase intergalactic medium \citep{SilkNusser10}. Star formation triggered by expanding radio galaxy lobes, especially those of the FR~II \citep{FR74} type, may explain the alignment between large scale optical emission and the axes of their relativistic plasma jets \citep[e.g.][]{McCarthy1987, Best1996, Eales1997, Croft2006}. Under simple orientation-based unified schemes, powerful radio galaxies and RLQs are drawn from the same parent population \citep[e.g.][]{Antonucci1993}. It is believed that the difference in these two populations in terms of their observed properties is the line of sight, which in the case of RLQs is closer to their emission axis \citep{Barthel1989}. Consequently, RLQs should also exhibit the alignment effect, albeit spatially unresolved due to the proximity to our line-of-sight.

There are three main physical interpretations for the alignment effect. The optical emission is proposed to be either: 1) dominated by light from young massive stars that formed when an expanding radio source (e.g. jets) collapse the dense gas providing added pressure \citep[e.g.][]{Bicknell2000, Fragile2004}, 2) scattered radiation from the AGN at the centre of the radio galaxy \citep{diSerego1989, Tadhunter1992}, or 3) due to Inverse Compton scattering of Cosmic Microwave Background (CMB) \citep{Daly1992}.

The first of these, where the radio source propagates into the interstellar and intergalactic medium can also produce shocks \citep[e.g.][]{Best2000}. Low-ionization emission lines could be produced by the ionizing photons in the cooling - shocked gas or, alternatively, by the warm clouds as they cool behind the shock. \cite{Solorzano2003} and \cite{Tilak2005} have confirmed that the radio lobes and hot spots are preferentially associated with lower ionization gas produced by precursor gas ionized by the shock being driven into the cloud by the deflected radio jet. However, as discussed by \cite{Bicknell2000}, following \cite{Elmegreen1978}, the shock ionized gas also tends to go hand-in-hand with the occurrence of star-formation as gravitational instability occurs on timescales comfortably within the dynamical timescale of the jet-cloud interaction. Therefore, although we cannot rule out purely low-ionization  shocked gas, it is unlikely that this is solely responsible for the excess [O{\sc ii}] emission.

\citet{Battye2009} compared the alignment effect for different radio-loud AGN at 1.4~GHz. They found that less-luminous RL AGN tend to align their radio emission axis with the minor axis of the starlight of the host. In contrast, no significant preference of radio-optical alignment is found for more powerful RL AGN. This study appears to be at least consistent with finding that there is higher [O{\sc{ii}}] equivalent width in RLQs with lower radio luminosity.

\subsection{Negative AGN feedback}

Semi-analytic models invoke AGN feedback to truncate star formation in massive galaxies \citep[e.g.][]{Croton2006, Bower2006}, often termed negative feedback. This feedback could take several forms such as the photoionizing and heating of the gas reservoirs in the host galaxy \citep[e.g.][]{Pawlik2009}, or the expulsion of significant amounts of gas through outbursts \citep[e.g.][]{Nesvadba2006}. The correlation between the formation of black holes and the star formation can be interpreted in two ways. In the first one, black hole accretion and star formation occur simultaneously because they are both fed from the same gas, brought to the center by gas-rich mergers and disc instabilities. Black hole accretion is then terminated when star formation has used up all the gas. In the second interpretation, star formation terminates when the black hole blows all the gas outside its host galaxy \citep[e.g.][]{Springel2005, Hopkins2006}.

An interesting observational result is that most massive galaxies since redshift $z\sim2$ appear to shut down star formation earlier than the less massive ones \citep[e.g.][]{Bundy2006, Bundy2008}. This phenomenon, known as `downsizing' \citep{Cowie1996}, coincides with the drop in the space density of quasars at recent epochs pointing to the possibility that black hole accretion and star formation are terminated together.

\subsection{A possible explanation}

The results of our analysis, and especially Figure~\ref{fig:sumup}, could tell a story about the star formation and [O{\sc{ii}}] emission in RLQs and RQQs. The break at optical luminosity/black hole mass versus [O{\sc{ii}}] equivalent width for the RLQs provides strong evidence for different evolution of star formation between the two quasar samples. It seems that when the black holes are less massive, presumably due to observing them earlier in their lifetimes, the two populations have significant differences in their [O{\sc{ii}}] emission, but they have the same distribution when they have accumulated the bulk of their final mass.

We propose that both of the populations start their activity due to a merger event or through an inflow of cold gas due to secular processes. As long as strong gravitational interactions play an important role in the production of super massive black holes and the quasar activity, this should be common for both radio loud and radio quasars. Depending on the pair of galaxies that were involved in the interaction and whether they can spin up a black hole, RLQs or RQQs could both be produced. For example, a high-mass/low-mass pair will produce a high mass black hole with low spin yielding a RQQ. On the other hand the merger of two high-mass galaxies would produce a high-mass, high-spin black hole and probably a RLQ. 

Physically, right after the halo merger, a torque is exerted in the host and gas is quickly funneled into the center of the newly-formed gravitational potential. In this phase, the black hole accretes its fuel very efficiently and the luminosity of the quasar increases. In order to explain the [O{\sc{ii}}] emission difference between the two populations, we need a mechanism which can trigger star formation in the case of RLQs. The obvious mechanism in this case are the jets. As the jets pass through the emission line region, the shock will provide additional ionizing photons, increasing the [O{\sc{ii}}] emission line. Radio jets are able to induce the formation of massive knots of bright young stars, which will disperse and fade over the lifetime of the quasar. This mechanism results in the tight alignment of the bright optical-UV continuum emission along the radio jet.

\begin{figure}
\includegraphics[scale=0.45]{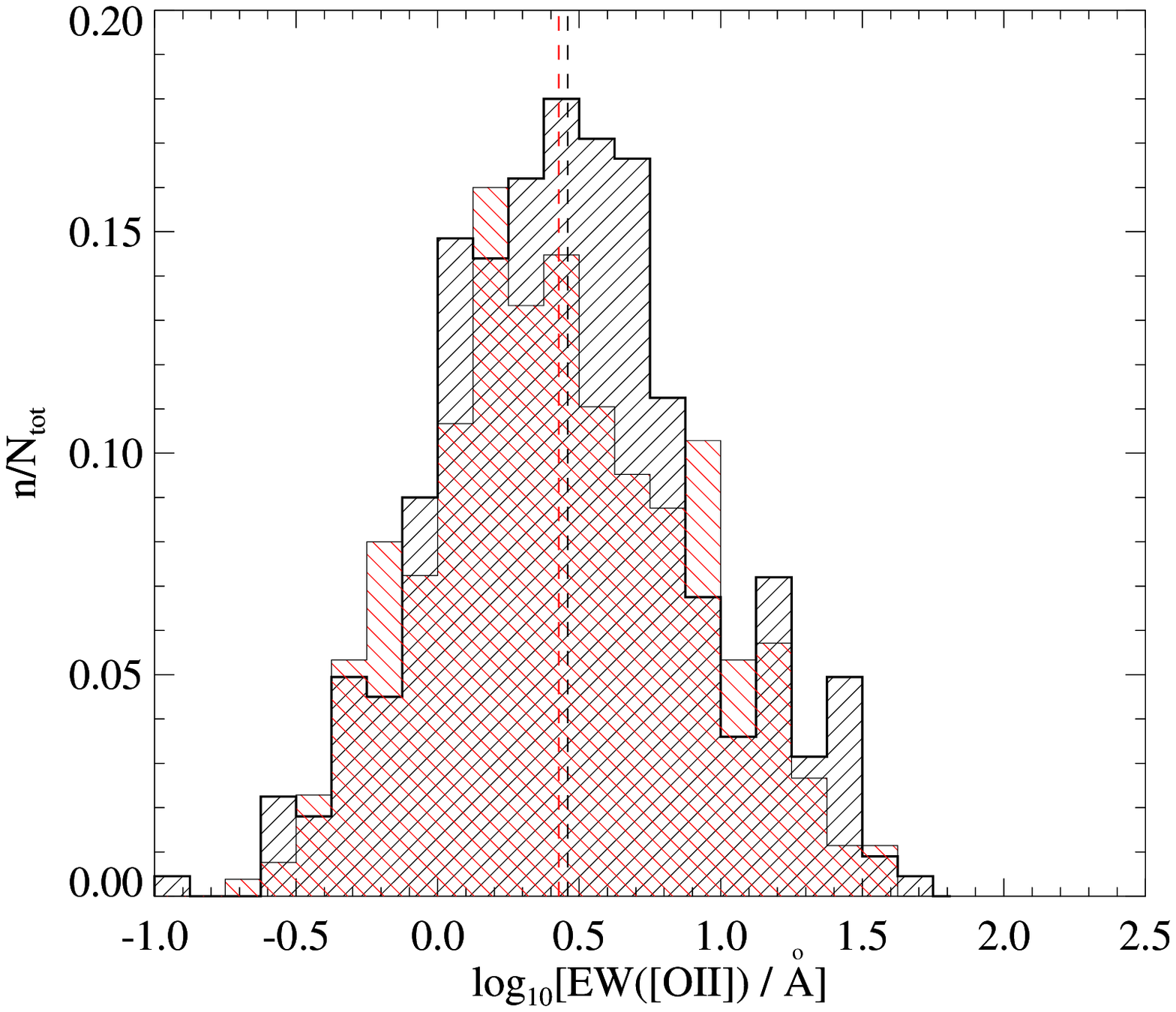}
\includegraphics[scale=0.45]{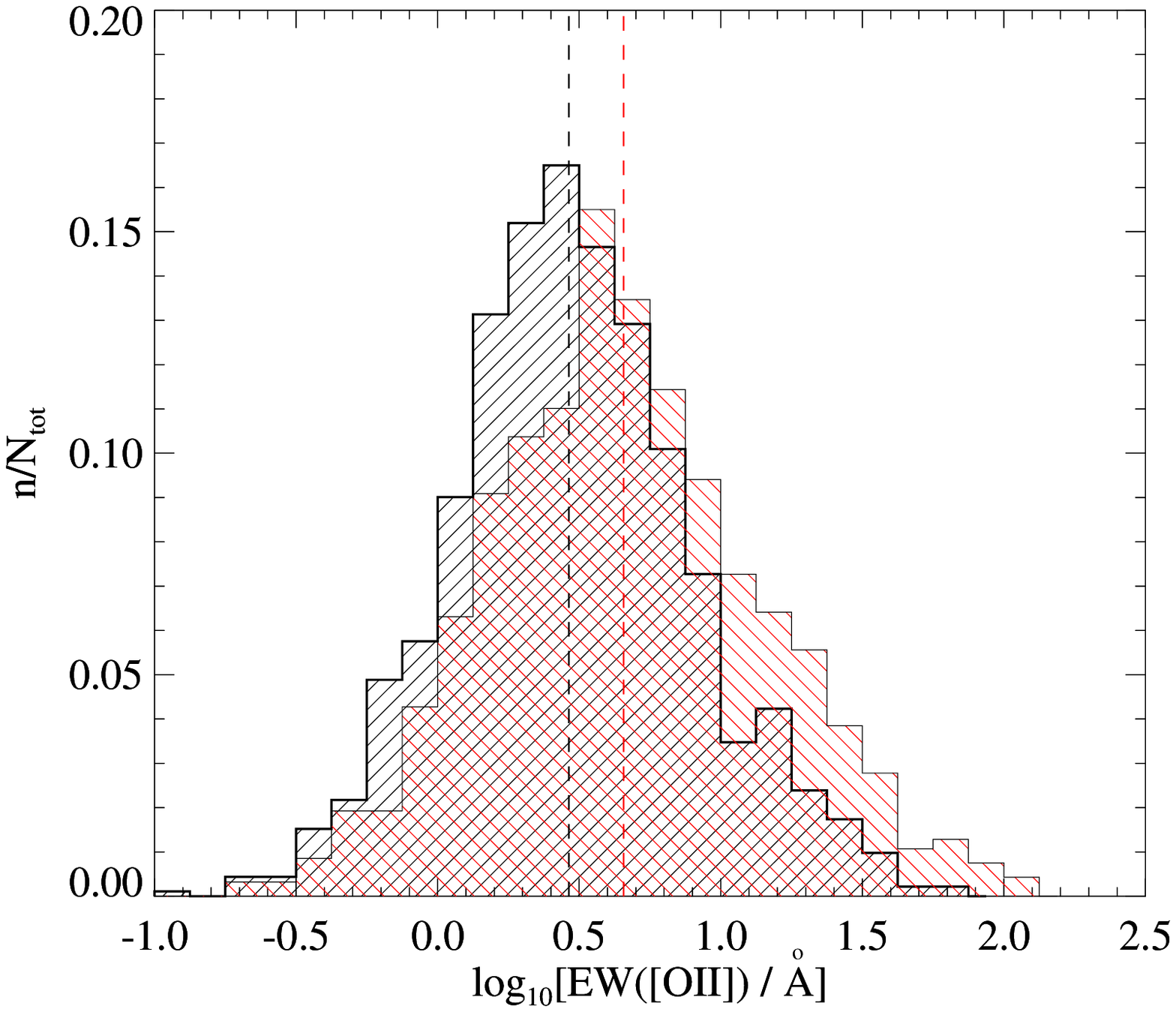}
\caption{Comparison of the ${\rm EW}($[O{\sc{ii}}]$)$ distribution for the sample of RLQs (red) and the matched sample of RQQs (black). The dashed lines shows the mean values for each population. The top panel shows the distribution for sources with $\log_{10}(L_{\rm opt} / {\rm W}) >38.6$ while the bottom for sources with $\log_{10}(L_{\rm opt} / $W$) \leq38.6$.}
\label{fig:last_comp}
\end{figure}

As the black hole mass increases, the jets are no longer able to keep the gas away from the black hole. A galaxy with a large bulge has been created and the new stars can feed the black hole. At this point the negative AGN feedback is stronger than the positive, the accretion rate and the black hole mass increase and as a consequence the luminosity of the quasar. The [O{\sc{ii}}] emission in RLQs decreases relative to the RQQs, and the RLQs and RQQs would have similar [O{\sc{ii}}] emission. For both of the populations the accretion can not last indefinitely, and it is quenched when the quasar is powerful enough to expel this gas or at least to balance its infall, thus inhibiting further growth.

In order to support the above explanation, in Figure~\ref{fig:last_comp} we compare the [O{\sc{ii}}] emission for the two populations for optical luminosities $\log_{10}(L_{\rm opt} / $W$)>38.6$ (top panel) and $\log_{10}(L_{\rm opt} / $W$)\leq38.6$ (bottom panel). The difference is obvious for the [O{\sc{ii}}] equivalent width. For $\log_{10}(L_{\rm opt} / $W$) >38.6$ the mean [O{\sc{ii}}] equivalent widths for the two populations are $<{\rm EW}($[O{\sc{ii}}]$)_{\rm RL}>=4.65 \pm 0.05$~\AA\ and $<{\rm EW}($[O{\sc{ii}}]$)_{\rm RQ}>=4.83\pm0.04$~\AA\ for RLQs and RQQs respectively. In contrast, for $\log_{10}(L_{\rm opt}  / $W$) \leq38.6$ the mean values are $<{\rm EW}($[O{\sc{ii}}]$)_{\rm RL}>=8.74\pm0.08$~\AA\ and $<{\rm EW}($[O{\sc{ii}}]$)_{\rm RQ}>=4.67\pm0.05$~\AA. The one-dimensional K-S test returns for $\log_{10}(L_{\rm opt} / $W$) >38.6$ a probability of $p=0.24$ and a test statistic of $D=0.08$ while for $\log_{10}(L_{\rm opt} / $W$)\leq38.6$ returns $p=1.04\times 10^{-16}$ and $D=0.18$.
For high optical luminosities the two populations are drawn from indistinguishable distributions and have the almost the same mean [O{\sc{ii}}] equivalent width, whereas for low optical luminosities RLQs have approximately twice the [O{\sc{ii}}] emission of RQQs. As we can see, in the case of RQQs there is no significant difference in [O{\sc ii}] emission for the two optical luminosity bins.

\section{Conclusions}

In this paper we have studied the relationship between the [O{\sc ii}] emission line properties of matched samples of RLQs and RQQs. The main result of our study is that RLQs have higher [O{\sc{ii}}] emission than RQQs. We find that the [O{\sc{ii}}] equivalent width is higher for RLQs at low optical luminosities. As the optical luminosity and also the black hole mass increase, the [O{\sc{ii}}] equivalent width for RLQs decreases while for RQQs it remains almost the same. Above an optical luminosity limit of $\log_{10}(L_{\rm opt} / $W$)=38.6$, the [O{\sc{ii}}] equivalent widths for the two samples are indistinguishable. Below this limit the $L_{\rm opt} - {\rm EW}($[O{\sc{ii}}]$)$ correlation changes for RLQs, while for RQQs it remains the same.

Based on these results, we believe that the main reason for this difference is the presence of powerful radio jets in RLQs that pass through the gas and delay the in-fall of the gas into the central supermassive black hole. At the same time they ionize the gas in the ISM (interstellar medium) and are able to induce star formation. When the black hole mass increases, along with the luminosity, due to the accretion of matter, the jet power is no longer enough to keep the gas away from the black hole and the  star formation decreases. At these high optical luminosities the two populations tend to have the same [O{\sc{ii}}] emission. However, we note that to obtain a full picture of the process occurring then a measurement of the environmental density would also be advantageous \citep[e.g.][]{Kauffmann2008, Falder2010, Falder2011}.

Even if [O{\sc{ii}}] emission is generally used as a star formation tracer, it is not one of the strongest indicators and our results may be biased due to the AGN emission especially at high optical luminosities \citep{Ho2005}, although we emphasize that our two populations have the same optical luminosity distribution. However, the fact that we show that the [O{\sc{ii}}] emission excess for the radio-loud population is present using either the [O{\sc iii}] luminosity or the optical luminosity as an AGN luminosity tracer, reinforces our suggestion that this excess is the result of higher star-formation activity in RLQs. A caveat to this is that some, possibly significant fraction of the [O{\sc ii}] emission could emanate from shocked gas within the radio lobes, although we note that such emission usually goes hand-in-hand with star formation. Our results therefore open the way for investigations in this direction using more secure star formation tracers such as far-infrared emission. This will be possible with surveys such as {\em Herschel}-ATLAS which cover a wide enough area to include a large number of the rarer RLQs.

\bsp

\label{lastpage}

\end{document}